   \newcommand{\bn}{\mathbf n}
 
\newcommand{\br}{\mathbf r}
 
\newcommand{\bx}{\mathbf x}

 \newcommand{\bF}{\mathbf F}

\newcommand{\dif}{{\rm d}}
\newcommand{\me}{{\rm e}}

\newcommand{\rM}{{\rm M}}

\newcommand{\ra}{{\rm a}}
\newcommand{\re}{{\rm e}}
\newcommand{\rs}{{\rm s}}
\newcommand{\rt}{{\rm t}}

\newcommand{\bnabla}{\boldsymbol \nabla}
\newcommand{\bOmega}{\boldsymbol \Omega}

\newcommand{\tb}[1]{\textcolor{blue}{#1}}
\newcommand{\tr}[1]{\textcolor{red}{#1}}
\newcommand{\tg}[1]{\textcolor{green}{#1}}
\newcommand{\ty}[1]{\textcolor{yellow}{#1}}

\newcommand\nobrkhyph{\mbox{-}}

\documentclass[twocol]{ametsoc}

\usepackage{tabularx}
\usepackage{algorithmic}
\usepackage{algorithm}
\usepackage{amsthm}

\journal{jas}

\bibpunct{(}{)}{;}{a}{}{,}

\title{Toward Cloud Tomography from Space using MISR and MODIS:\\
The Physics of Image Formation for Opaque Convective Clouds}

\authors{Anthony B. Davis
\correspondingauthor{NASA JPL, 4800 Oak Grove Drive, MS 233-200, Pasadena, CA.} 
}

\affiliation{Jet Propulsion Laboratory, California Institute of Technology, Pasadena, California, USA}

\extraauthor{Linda Forster
\thanks{Also: Ludwig-Maximilians-Universit{\"a}t, Munich, Germany.}
}
\extraaffil{Jet Propulsion Laboratory, California Institute of Technology, Pasadena, California, USA}

\extraauthor{David J. Diner}
\extraaffil{Jet Propulsion Laboratory, California Institute of Technology, Pasadena, California, USA}

\extraauthor{Bernhard Mayer}
\extraaffil{Ludwig-Maximilians-Universit{\"a}t, Munich, Germany}

\email{Anthony.B.Davis@jpl.nasa.gov}

\abstract{
3D convective cloud images form via two intertwined radiative diffusion processes.
Sunlight starts in the anti-solar direction and ends in toward-sensor ones, but repeated forward-peaked scattering smears the well-collimated beams across \emph{direction} space.  
This loss of directional memory in the cloud's ``outer shell'' (OS) is modeled as a random walk (RW) on the sphere.  
We show that, for typical cloud phase functions, 5 or 6 scatterings suffice for severe degrading of directionality.
Simultaneously, a RW unfolds in \emph{standard} 3D space where steps are angularly-correlated, hence a drift in the original direction and an associated lateral dispersion. 
Any distinctive cloud image ``feature'' originates in the OS, and the shallower the better.
That is also why we previously found that the optical depth of the ``veiled core'' (VC) is $\approx$5.
Significant amounts of sunlight thus arrive at the VC as a diffuse irradiance, and leave it even more isotropic.  
The diffusion limit of 3D radiative transfer (RT) is therefore valid inside the VC.  
Consequently, the underlying RW in the VC unfolds in 3D space, now with isotropic steps since extinction is scaled back to account for forward scattering.
We show that the VC optical thickness controls cloud-scale brightness contrast between the illuminated and self-shaded sides of the cloud.
Full cloud image formation thus involves diffusions (i.e., RWs) in both Euclidian and spherical/direction spaces.
3D cloud tomography based on MISR and MODIS multi-angle/-spectral data is an emerging technique in passive VNIR-SWIR sensing that will make judicious use this spatial separation of RT regimes to accelerate forward modeling without significant loss in accuracy.
}

\begin{document}

\maketitle


%








\section{Introduction \& Outline}
\label{sec:intro}

Our poor understanding of clouds has been identified as a major source of uncertainty in the prediction of future climate scenarios with Earth-system models used in future climate prediction \citep{IPCC_AR5}.
This unfortunate state-of-affairs concerns the radiative properties of all clouds types as well as their role in the hydrological cycle, including their complex interactions with natural and anthropogenic aerosols.
Due to their propensity for producing precipitation, clouds that form in convective dynamical regimes, both shallow and deep, over both water and land, are particularly problematic.

Normally, active (i.e., lidar and radar) and passive remote sensing would be a reliable source of knowledge about clouds and precipitation, with the later modality providing the spatial coverage and the former the vertical structure.  
Clearly, no single observational technology reveals all we need to know.
NASA is therefore proactively selecting, under a fixed budget and in consultation with the scientific community, the optimal sensor suite for its next generation of satellites, in pursuit of science goals spelled out in the 2017 Decadal Survey (DS17) for Earth Science \citep{DS2017}.

The DS17 ``Aerosol/Cloud-Convection-Precipitation'' track is confronted with conflicting needs to maintain a robust program-of-record (POR) and for innovation.
Operational cloud sensing in reflected solar light is currently limited by its assumption of plane-parallel cloud geometry, hence 1D radiative transfer (RT) to predict visible/near-IR (VNIR) and shortwave-IR (SWIR) observations \citep{platnick2017}.
Biases ensue in the retrieved cloud properties (cloud optical thickness, effective particle size), even under the best circumstances, i.e., extended stratiform cloud layers that approximate the assumed plane-parallel geometric model \citep{NK_bispectral_90}.  
Due to their inherently 3D nature, vertically-developed convectively-driven clouds are all but forsaken in the sense that retrievals are performed on them but rarely pass tests for reliability \citep{cho2015}.

\begin{figure}
\begin{center}
\includegraphics[width=3.14in]{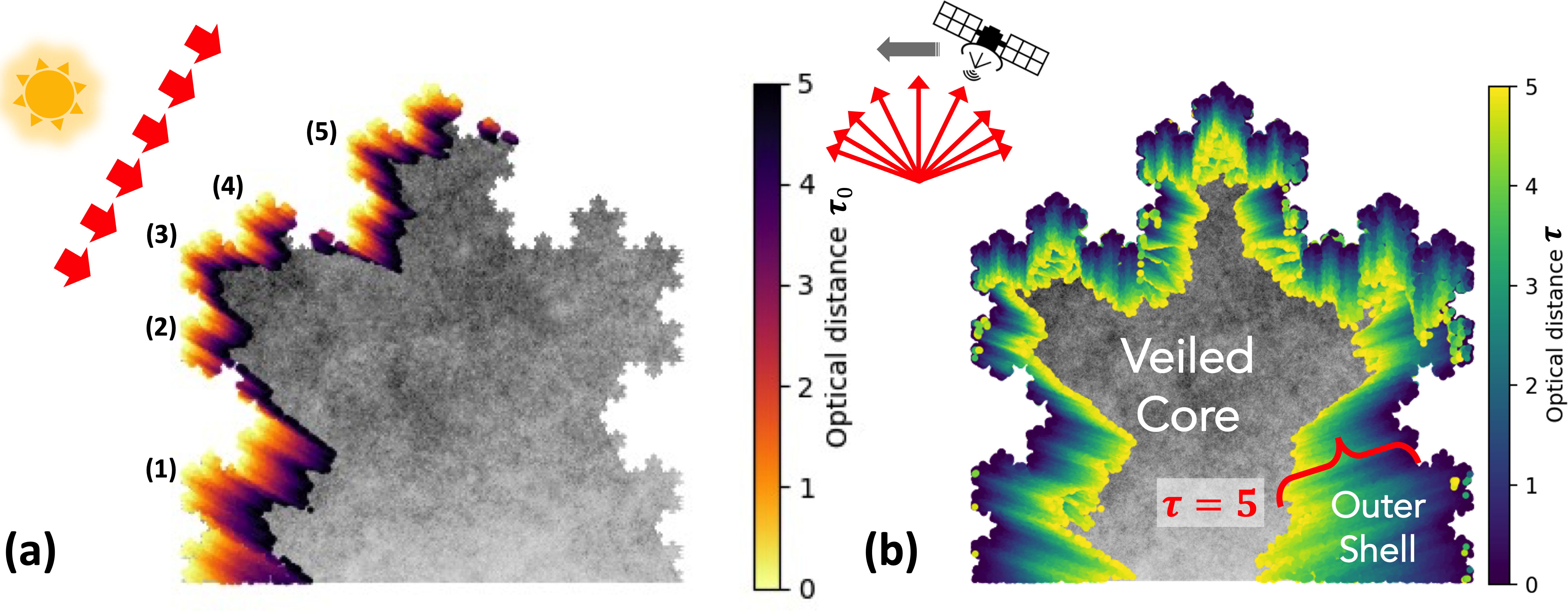}
\end{center}
\caption{
OS and VC of a 2D toy cloud model described by \cite{Forster_etal21} (e-supplement) based on the Koch fractal for its outer shape, and fractional Brownian terrain for its inner structure, tuned to mimic turbulence.
There is also a cloud-scale vertical gradient to capture how an adiabatic parcel-model would distribute liquid water content (LWC) vertically.
Optical thickness along the central column is set to 40. 
{\bf (a)} This version illustrates solar illumination for $\theta_0$ = 60$^\circ$. 
Numbered asperities are discussed further on (Fig.~\ref{fig:Koch_R_vs_T} in \S\ref{sec:VC_processes}b).
{\bf (b)} This version visualizes the whole VC by accounting for all 9 of MISR's viewing angles: 0$^\circ$, $\pm$26.1$^\circ$, $\pm$45.6$^\circ$, $\pm$60.0$^\circ$, and $\pm$70.5$^\circ$, respectively, for An, Aa/f, Ba/f, Ca/f, and Da/f cameras.
Any point within the colored areas is in the OS, while the VC is the grey zone.  
}
\label{fig:Koch_in+out_VC}
\end{figure}

We believe that, to break out of the paradigm of pixel-by-pixel 1D RT-based retrievals that operational passive cloud sensing is committed to---now because of additional pressures of POR continuity---we need to improve our fundamental understanding of how images of optically thick 3D clouds are formed.
Out of this quest, a tale emerges where two intertwined radiative diffusion processes unfold in the cloud's ``veiled core'' (VC) and ``outer shell'' (OS).  
In essence, the OS is the radiative equivalent of a boundary layer as defined in fluid dynamics, that is, where the presence of boundary sources and sinks of solar radiation are strongly felt.
In the OS, radiation-matter interactions thus shape the structure of the radiance field, including multi-angle imaging by remote sensors.  
The VC is defined empirically in a previous study \citep{Forster_etal21} as the potentially large inner region of the cloud where small-scale details of the extinction field have a negligible impact on the remotely-observed escaping radiance fields, i.e., multi-view images. 
A ``negligible'' effect is defined here as not exceeding the level of sensor noise.
Figure~\ref{fig:Koch_in+out_VC} uses a fractal cloud model to visualize the VC and OS.
Although Fig.~\ref{fig:Koch_in+out_VC} shows a sharp boundary between the OS and VC, the physically-correct definition presented further on leads to a more gradual transition.

Cloud tomography (CT) using fused multi-angle/multi-spectral imaging from the likes of the Multi-angle Imaging SpectroRadiometer \citep[MISR,][]{Diner_etal98} and MODerate resolution Imaging Spectrometer \citep[MODIS,][]{Salomonson_etal02} on Terra is a much-needed breakthrough in passive VNIR-SWIR cloud remote sensing from space.
CT has been demonstrated so far \citep{Levis_etal15,Levis_etal17,Levis_etal20} only with synthetic and real airborne imaging sensor data where image pixels and grid voxels are relatively small, hence optically thin and arguably homogeneous, which is consistent with assumptions in the forward model at the core of the retrieval, namely, SHDOM \citep{evans1998spherical}.\footnote{
An alternative 3D RT-based approach to CT has been formulated mathematically \citep{martin2014adjoint} and demonstrated \citep{martin2018demonstration}, but so far only on a 2D transect through an LES cloud field.}
Moreover, the clouds subjected to CT so far are relatively small, so no particular attention was paid to the OS/VC partition.
Specifically, data from the Airborne Multi-Spectral Imaging Polarimeter (AirMSPI) \citep{diner2013} was used, with $\sim$20~m pixels, after rigorous uncertainty quantification (UQ) of the retrieval methodology using synthetic data from Large-Eddy Simulation (LES) of vertically-developed 3D clouds and high-fidelity 3D RT \citep{Levis_etal15}.
Again, key to the success of the CT demonstration is that voxels and pixels are comparable in scale at a few 10s of meters, and can reasonably be assumed optically thin and internally uniform.

\begin{figure}[htbp]
\begin{center}
\includegraphics[scale=0.7]{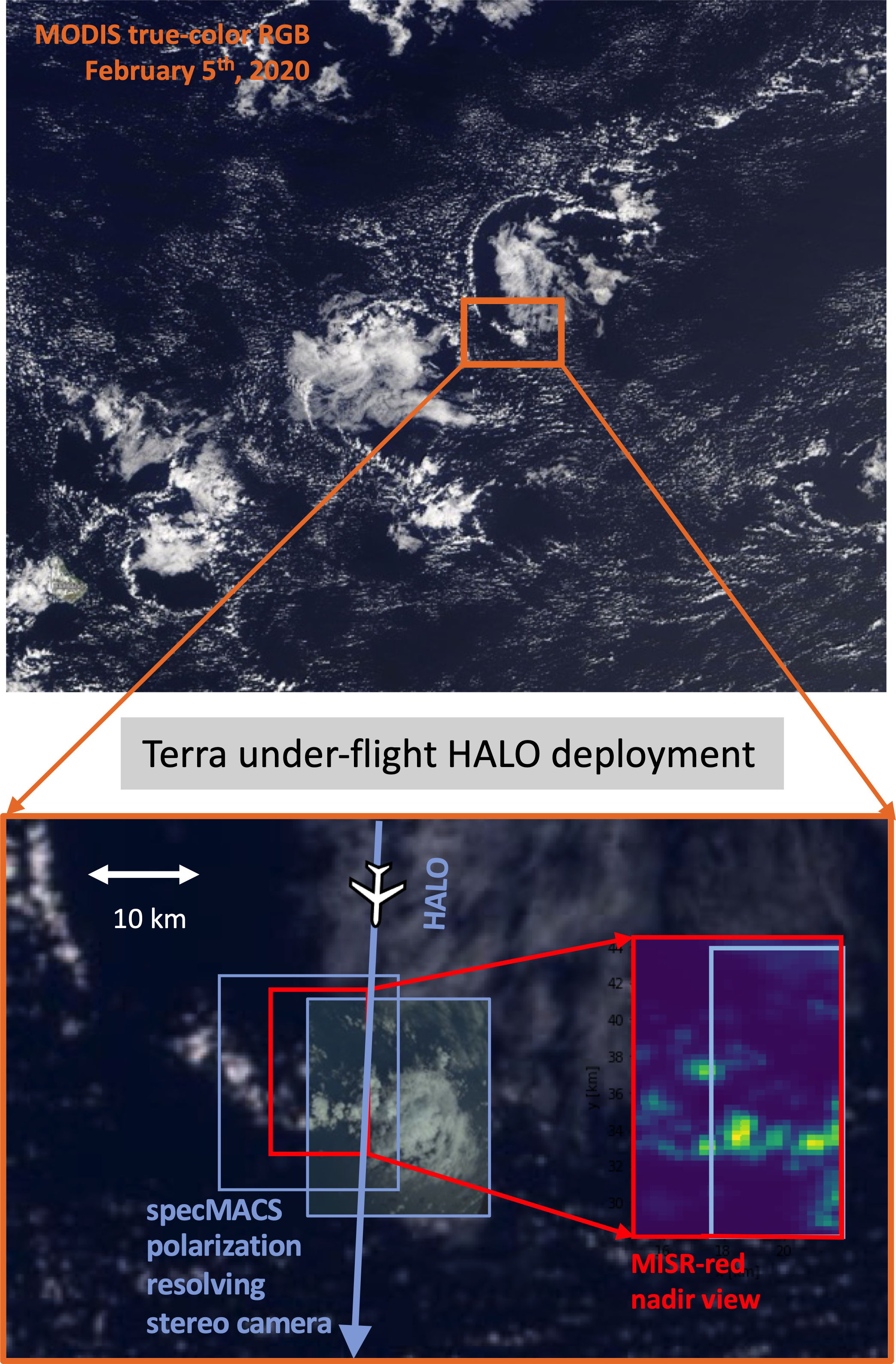}
\caption{
VIS imagery from MODIS and MISR and the airborne specMACS polarization resolving stereo camera captured on Feb 5$^\text{th}$, 2020, while the DLR HALO (High Altitude and LOng range research) aircraft was under-flying the Terra satellite during the EUREC$^4$\!A field campaign. 
Opaque 3D clouds of various sizes are observed by all 3 sensors and are thus candidates for reconstruction using CT, with retrievals from the fine-scale airborne imagery offering a form of validation for CT methods now under development for coarse-scale satellite data.
}
\label{fig:EUREC4A}
\end{center}
\end{figure}

The situation changes radically for above-mentioned space-based sensors where pixels are 14 (MISR-red) to 25 (MODIS-SWIR) times larger than those of AirMSPI.
The cloudy atmospheric columns defined by MISR or MODIS pixels will almost surely be optically thick in \emph{horizontal} directions as well as internally variable.  
These simple facts are problematic for current SHDOM-based CT codes because SHDOM assumes grid cells to be uniform and optically thin in all directions.
Figure~\ref{fig:EUREC4A} (top panel) displays a 250~m resolution MODIS ``true color'' image captured during the EUREC$^4$A (ElUcidating the Role of Clouds-Circulation Coupling in ClimAte) field campaign \citep{Bony_etal17,Stevens_etal21}.
The bottom panel in Fig.~\ref{fig:EUREC4A} shows a subset of the MODIS data in the top panel, along with colocated data from MISR (red channel, 275~m resolution).
Also shown is collocated imagery from the spectrometer of the Munich Aerosol Cloud Scanner \citep[specMACS,][]{Ewald_etal16}, rendered at $\approx$20~m resolution, that was onboard the DLR High Altitude and Long Range Aircraft (HALO) during the Terra under-flight.
This deliberate deployment during EUREC$^4$A will create an opportunity for validation of CT using relatively coarse-scale data from space-based sensors, which is currently under development.

Fortunately for future development of CT from sensors in space, we will show herein that there is a clear spatial separation of transport regimes:
\begin{itemize}
\item
\emph{standard 3D RT} in the OS controls the gradual blurring of internal cloud structure in the imagery;
\item
\emph{radiative diffusion} in the VC controls the cloud-scale gradient in brightness between the illuminated and self-shaded sides of the cloud.
\end{itemize}
The formulation of space-based CT as a large ill-posed inverse problem will be informed by these insights, and new hybrid forward models for CT will make judicious use of them.
Even the initialization of necessarily-iterative CT schemes can be expedited using the diffusion theoretical prediction for the contrast ratio of mean radiances for the sunny and shady sides of vertically-developed convective clouds \citep[cf.][]{davis2002}.

The paper is organized as follows.
In upcoming Section~\ref{sec:OS_processes}, we investigate quantitatively radiative processes that unfold in the OS, with the theoretical background being covered in e\nobrkhyph{}Supplement~``A'' (hereafter referred to as ``Appendix~A'').
Section~\ref{sec:VC_processes} addresses radiation transport across the VC, with the underlying theory being covered in e\nobrkhyph{}Supplement~``B'' (hereafter referred to as ``Appendix~B'').
In Section~\ref{sec:OS_VC_interaction}, we examine how the OS and VC interact radiatively and, ultimately, how entire cloud images are formed.
We summarize and discuss our findings in Section~\ref{sec:discuss} and apply them to a case study in Section~\ref{sec:Case_study}.
We draw our conclusions in Section~\ref{sec:concl} and describe future research in support of cloud CT using merged MISR and MODIS data from NASA's flagship Terra platform.

\section{Directional smearing and spatial drift with lateral dispersion in the OS}
\label{sec:OS_processes}

\subsection{Connection between image features and cloud structures goes from sharp to fuzzy}

\begin{figure}[htbp]
\begin{center}
\includegraphics[width=3.14in]{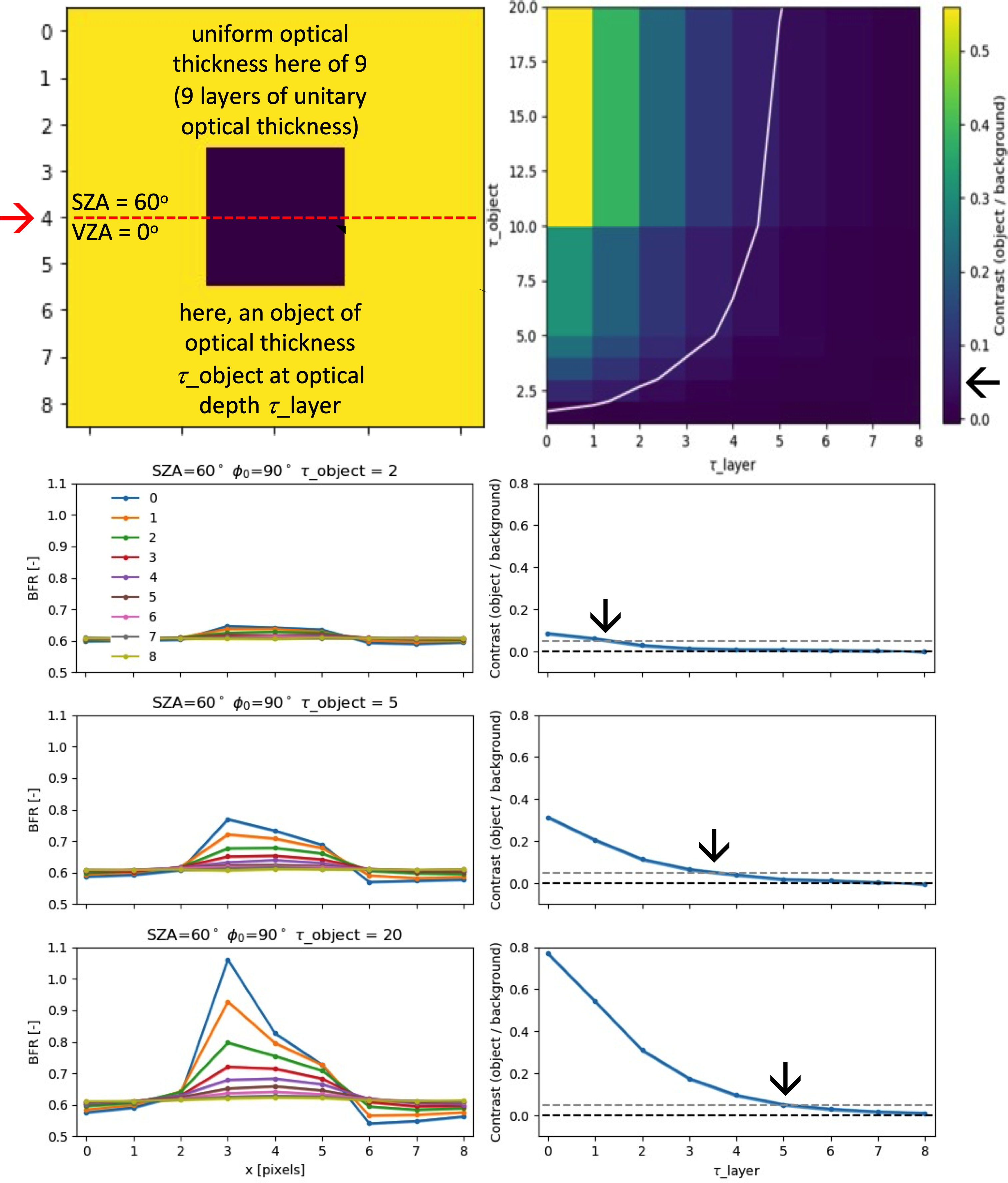}
\end{center}
\caption{
The upper right panel schematic depicts an embedded ``object'' in an otherwise uniform plane-parallel medium of optical thickness 9: a 3x3x1 voxel disturbance viewed from above; SZA is 60$^\circ$ coming from the left, as indicated with the dashed line.
The three lower left-hand panels show the resulting 3 nadir pixel-level radiances across the middle of the object (dashed line) along with 3 more pixels on either side of it, hence 9 pixels in all.
From top to bottom, 3 selected optical thicknesses of the object are used: $\tau_\text{object}$ = 2, 5, and 20, in an environment where the optical thickness of a single layer is unity.
All radiances are computed to high precision, with $\approx$0.5\% numerical error.
The image transects are displayed for optical depths from the upper boundary down to the top of the object (denoted $\tau_\text{layer}$) varying between 0 to 8 (color coding in first panel applies to all).
Where perceptible, the width of the lines captures the Monte Carlo noise level (2 standard deviations).
The three lower right-hand panels show the contrast ratio defined as the maximum-to-minimum radiance difference (denoted as ``feature'') divided by to the mean radiance along the transect in the left panel (denoted as ``background'').
This ratio is a metric for the strength of the apparent image feature, and is plotted as a function of $\tau_\text{layer}$.
The horizontal grey line indicates the highlighted 5\% contrast level at which the feature barely emerges from the radiometric sensor noise (i.e., $\approx$3\% for each radiance in the difference).
Finally, the upper right panel compiles the contrast ratio as a function of both $\tau_\text{layer}$ and $\tau_\text{object}$, with the critical 5\% contrast level highlighted in white.
}
\label{fig:emerging_features}
\end{figure}

An important concept at the core of CT is the causal connection between the 2D spatial variability of cloud images and the 3D spatial structure of the cloud.
Figure~\ref{fig:emerging_features} displays a series of high-precision numerical experiments using the MYSTIC 3D RT solver \citep{mayer2009,buras2010} that shows how an image feature is associated with a simple internal cloud structure.
A geometrically plane-parallel cloud is divided into 9 equally-spaced layers, each of optical thickness unity.
There is a Lambertian surface with an albedo of 2/3 at the lower boundary while, at the upper boundary, the solar beam impinges on the medium at solar zenith angle (SZA) 60$^\circ$.
The top 5 or 6 layers can be thought of as the OS, and the lower ones along with the partially reflective surface as the VC.
The domain is divided into 9$\times$9 = 81 MISR-like pixels.
An ``object'' is embedded in the cloudy medium that covers the central 3$\times$3 = 9 pixels, and has the same \emph{physical} thickness as the layer (see upper right panel of Fig.~\ref{fig:emerging_features}), but its \emph{optical} thickness exceeds that of the background by factors of 2, 5 and 20, from top to bottom.

As expected, we see in Fig.~\ref{fig:emerging_features} that the stronger the internal extinction gradients, the stronger the resulting image feature.
Moreover, the presence of the object in the cloudy medium is easily detectable in the observations if it is at sufficiently shallow optical depth.
However, the image feature that the object creates eventually blends into the sensor noise as it sinks to ever larger optical depths at a rate that depends on the ``strength'' of the cloud structure, i.e., how opaque (or tenuous) it is compared to the background.
Objects of optical thickness $\tau_\text{object}$ up to 20$\times$ the background create image gradients that vanish into the radiometric noise at an optical depth $\lesssim$5.
As expected, even the most opaque objects can go undetected after they reach the OS/VC interface at $\tau_\text{layer} \approx 5$, and sink further into the VC.
The upper right panel in Fig.~\ref{fig:emerging_features} regroups the results in the right-hand plots of Fig.~\ref{fig:emerging_features} as well as intermediate cases in a 2D density plot in ($\tau_\text{layer},\tau_\text{object}$) space.
The feature ``emergence'' threshold of 0.05 for the adopted contrast ratio (between the max-to-min radiance difference normalized by the image transect mean) is highlighted.
To be detectable as an image feature, the object's parameters must be to the left of the white line: as anticipated, opaque and/or shallow objects generate observable features.
This experiment reinforces the choice of $\approx$5 as the threshold optical thickness of the OS.

In this case of oblique illumination (SZA = 60$^\circ$), we also see that change in radiance value across pixels is invariably steeper on the illuminated side of the object than on its shaded side, even when the object is not particularly opaque.
This is because radiation flows across more pixels (i.e., cloud columns) when the light is more in the diffuse field and less in the direct beam---a phenomenon known as radiative smoothing \citep{Marshak_etal1995,Davis_etal1997}.
That said, all gradients are naturally flattened as the object goes deeper into the OS and eventually the VC.

At any rate, the sharper pixel-to-pixel gradients are more conducive to accurate feature detection and tracking in a horizontal wind and/or stereo height determination.
The conventional definition of ``cloud top'' in stereographic CTH retrievals \citep[e.g.,][]{marchand2001,HorvathDavies2004} is $\approx$1 optical depth below the actual cloud top, as defined by the highest occurrence of any condensed water particles, which can be viewed as the ``microphysical'' definition of cloud top. 
Figure~\ref{fig:emerging_features} shows that there is in fact a range of possible optical depths below the microphysical cloud top.
If we could determine for each detected feature a robust metric for its ``strength,'' then level curves in a 2D plot such as the upper right panel of Fig.~\ref{fig:emerging_features} would help to constrain the actual optical depth of the physical heterogeneity that caused the image feature.
From there, we could learn how to relate optical to physical depth, and thus refine the determination of actual CTH and the height assignment of the horizontal wind retrieval.

{\bf The OS of any cloud, cumuliform or stratiform, is therefore a region of gradual transition where sunlight goes from streaming ballistically outside the cloud to wandering randomly in space when it reaches the VC.}
That means that we have to pay attention to how directional properties interact with spatial counterparts.
From the remote sensing standpoint, the OS is where we will find the internal cloud structures that lead to ``features'' in the satellite or airborne imagery, e.g., for cloud top height (CTH) determination using stereographic methods.
From the cloud tomography perspective, the OS is where we should focus the bulk of the computational effort to reconstruct 3D cloud structure since that is where the spatial information encoded in the multi-angle imagery is connected with inherent 3D cloud structure.
In short, this is the realm of RT per se in the most 3D sense: details of the phase function (PF) matter, as do those of the spatial structure of the extinction field.
Notwithstanding, we show in the following that insights into the gradual transition are gained by focussing primarily on the angular distribution of the radiance field in the case of unbounded uniform 3D media.

\subsection{Angular distribution of radiance in the OS from a random walk on the 2D sphere} 

\begin{figure}
\begin{center}
\includegraphics[width=3.14in]{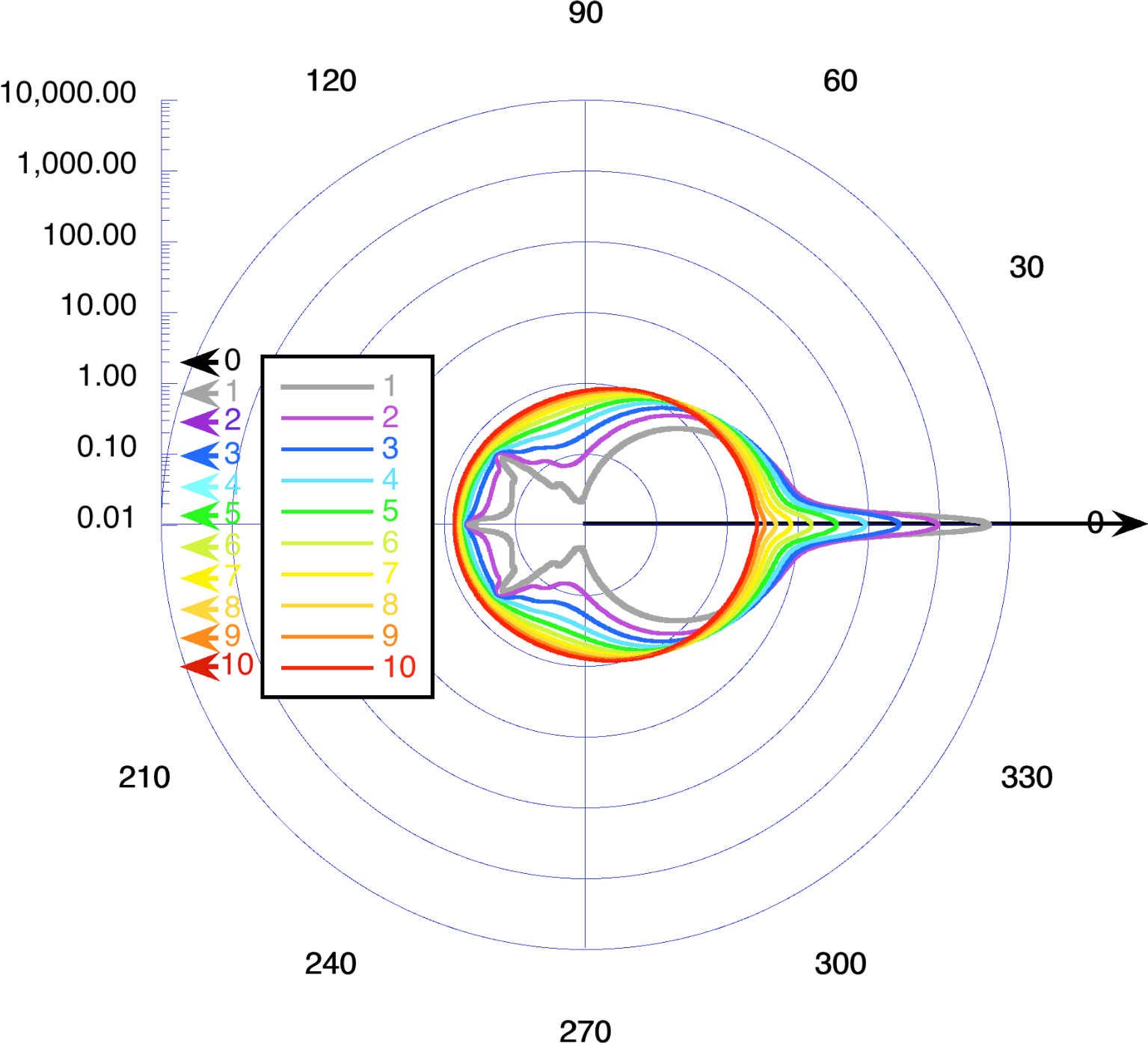}
\end{center}
\caption{
A typical liquid water cloud phase function convolved $N$ times with itself for $N$ = 0, 1,..., 10, on a log-scale.
Specifically, we assume $r_\me$ = 10~$\mu$m, $v_\me$ = 0.1 in a Gamma particle size distribution (PSD) from \citep{cahalan2005}, for $\lambda$ = 670~nm (MISR's red channel).
Outcomes for increasing $N$ are color-coded: $N = 0$, black; $N = 1$, gray; $N \ge 2$, rainbow colors with increasing wavelengths. 
Thus, ballistic motion is in black, singly-scattered radiance is in gray, multiply-scattered radiance is in color, going from cool to warm.
We clearly see the gradual erasing of identifiable single-scattering features such as the strong forward peak at $\theta = 0$, the back-scatter/glory peak at $\theta = 180^\circ$, as well as the primary and secondary rainbows at $\theta \approx 142^\circ$ and $126^\circ$, respectively.
}
\label{fig:PF_convolve}
\end{figure}

Figure~\ref{fig:PF_convolve} shows the rotationally invariant distribution of radiance as a function of polar angle $\theta$ after $N$ scatterings, $0 \le N \le 10$.
In App.~A, we are reminded that, at every scattering event, the prevailing angular distribution of radiance is convolved (in the spherical sense) with the PF.
Therefore, as far as direction of propagation is concerned, the radiance field is represented by successive convolutions of the PF with itself.
At $N = 0$, the light is confined to a perfectly collimated beam.
At $N = 1$, we simply have an image of the adopted cloud droplet PF, with over 5~orders-of-magnitude in range.
After only $N = 5$ scatterings, the range is drastically reduced to only one order-of-magnitude.
At $N = 10$, the field is nearly isotropic.

Following the formulation of RT in App.~A, we can recast this radiance field evolution as a discrete-time random walk (RW) on the unit sphere, i.e., the transport direction sub-space mapped out with polar coordinates $(\theta,\phi)$, where $\theta \in [0,\pi]$ and $\phi \in [0,2\pi)$.
Let $p(\mu_\rs)$ be the PF, where $\mu_\rs = \cos\theta_\rs$, with $\theta_\rs$ denoting the scattering angle.
We normalize $p(\mu_\rs)$ such that $2\pi \int_{-1}^{+1} p(\mu_\rs) \dif\mu_\rs$ is unity.
As usual in RW theory, all the ``particles'' are released at the North pole ($\theta = 0$) in uniformly random azimuthal directions.
At time $N = 1$, the particles are thus distributed across the sphere uniformly in azimuthal angle, as a reflection of the rotational symmetry of the PF.
In sharp contrast with azimuthal angle $\phi$, particles are now distributed very unevenly in polar angle $\theta$, specifically, according to the highly variable PF in Fig.~\ref{fig:PF_convolve}.
At time $N = 1$, the polar angle is simply $\theta = \cos^{-1}\mu_\rs$.

Of special interest in RW theory is the mean step size, a.k.a. mean-free-path (MFP), 
\begin{eqnarray}
\langle \theta_\rs \rangle &=& 2\pi \int_0^\pi \theta_\rs p(\cos\theta_\rs) \sin\theta_\rs \dif\theta_\rs \nonumber \\
                           &=& 2\pi \int_{-1}^{+1} \cos^{-1}\mu_\rs p(\mu_\rs) \dif\mu_\rs.
\label{eq:mean_theta_s}
\end{eqnarray}
For the PF in Fig.~\ref{fig:PF_convolve}, we have $\langle \theta_\rs \rangle$ = 18.9$^\circ$.
For comparison, the full-width-half-max (FWHM) of the forward peak is only $\approx$1.7$^\circ$, and one half of the scattering events are in the cone defined by $\theta \lesssim 5.5^\circ$.
In RT theory, we are more familiar with is the PF's asymmetry factor
\begin{equation}
g = \langle \mu_\rs \rangle = 2\pi \int_{-1}^{+1} \mu_\rs p(\mu_\rs) \dif\mu_\rs, 
\label{eq:g_factor}
\end{equation}
which is 0.86 for the PF in Fig.~\ref{fig:PF_convolve}.
An alternative metric for the RW step size is therefore $\cos^{-1}g$ = 30.5$^\circ$ for the PF in Fig.~\ref{fig:PF_convolve}.


So much for $N = 1$.
After $N = 5$ scatterings, the \emph{effective} asymmetry factor of the iterated PF is $g^5 \approx 0.47$ from App.~A, Eq.~(A2), and the associated polar angle has already more than doubled on average to 61.7$^\circ$.
After $N = 10$ scatterings, we have $g^{10} \approx 0.22$ and we are 77$^\circ$ away from the North pole.
In other words, memory of the original direction of propagation is all but lost, almost as it would be after a single isotropic scattering.
In App.~A, we show that for this RW on the sphere, positional correlations decay exponentially.
Technically speaking, it is a Markov process with \emph{short-term} memory.
Fundamentally, this is because the sphere is a finite manifold in 3D space.
{\bf The characteristic (a.k.a. ``e-folding'') time for the $g^n$ series is}
\begin{equation}
n_\text{smear} = -1/\log g.
\label{eq:n_smear}
\end{equation}
If $g \lesssim 1$, then $n_\text{smear} \approx g/(1-g)$; see App.~A, Eq.~(A7).
Numerically, these two estimators lead to $n_\text{smear}$ = 6.7 and $\approx$ 6.2, respectively, for $g$ = 0.86.

\subsection{Thickness of the OS, from longitudinal drift due to forward scattering}

What are the spatial ramifications of the RW on the sphere for the associated discrete-time RW in 3D space?
The strong forward peak of the PF forces the position of the particle to stay, on average, close to the axis defined by its original direction of propagation, say, the $z$-axis.
We show in App.~A that after $N$ steps, the mean position along the $z$-axis, denoted $\langle z_N \rangle$, is $\ell \, (1-g^{N+1})/(1-g)$, where $\ell$ is the MFP in 3D space.
In a uniform optical medium, $\ell = 1/\sigma$, where $\sigma$ is the extinction coefficient.
In optical distance, that systematic drift is thus given by 
\begin{equation}
\langle \tau_N \rangle = \sigma\langle z_N \rangle = \frac{\langle z_N \rangle}{\ell} = \frac{1-g^{N+1}}{1-g}.
\label{eq:mean_z_drift}
\end{equation}
How far has the 3D space RW drifted after $N = n_\text{smear}$ steps?
For a typical cloud droplet PF with $g$ = 0.86, we estimated that $n_\text{smear} \approx 6.5$, which yields $\langle \tau_{n_\text{smear}} \rangle \approx 4.8$.

The proximity of this optical distance with the optical depth of the OS/VC interface determined empirically by \cite{Forster_etal21}, namely, $\approx$5 is not coincidental.
Those authors define the 3D cloud's VC as the inner region where the extinction field can be rearranged in physically plausible ways, but the resulting changes in the remotely measured radiances are commensurate with sensor noise.
Approaching the VC from the opposite direction, one can ask:
{\bf How \emph{optically} deep can a reasonably strong fluctuation in the cloud's 3D extinction field be, and still be detectable in the outgoing 2D radiance field (i.e., an image of the cloud)?}

This is, in essence, a question about visibility through cloudy air.
It is also a key question in CT in the same way that the VC is, because any extinction field structure deeper than this threshold has little impact on the outgoing radiance fields, and is not worth trying too hard to retrieve.
The numerical experiment summarized in Fig.~\ref{fig:emerging_features} investigated this question empirically, but now we can address it theoretically with the above general results in RT.  

Considering only directionality of light leaving the region of fluctuation in the extinction field in the direction of a remote imaging sensor, it is more-and-more dispersed as the object sinks deeper into the cloudy medium.
This means there is less-and-less light heading toward the pixels that would sharply image that region in the absence of cloud. 
So, apart from being extinguished, the remaining light is spread over more-and-more pixels at the focal plane.
We venture that the threshold optical depth is roughly $\langle \tau_N \rangle$ in (\ref{eq:mean_z_drift}), which works out to be $\approx$5 for $N = n_\text{smear}$ in (\ref{eq:n_smear}) when $g$ = 0.86.
Now, a perfect sensor would always register some difference as the optical distance to the region of interest increases, but real sensors have noise floors that determine change detectability limits.
Typical signal-to-noise ratios (SNRs) for good VNIR/SWIR sensors are in the 100s.
However, direct or near-direct transmission across an optical distance of $\langle \tau_{n_\text{smear}} \rangle \approx 5$ is $\me^{-5} \approx 6.7\,10^{-3}$, which would be approaching the noise level (assuming a typical cloud reflectivity does not saturate the sensor).

This prediction should be quite robust since two extreme limits lead to logical answers.
First, in the theoretical limit of no scattering at all ($g \to 1$), then $n_\text{smear} \to \infty$ and $\langle \tau_{n_\text{smear}} \rangle \to n_\text{smear}+1 = \infty$ in (\ref{eq:mean_z_drift}), which is logical (at least in the absence of absorption, i.e., pure unattenuated ballistic propagation).
Second, in the limit of isotropic scattering ($g \to 0^+$), then $n_\text{smear} \to 0$ and $\langle \tau_{n_\text{smear}} \rangle \to n_\text{smear}+1 = 1$, which is also logical since there is no forward scattering trend for light to get any deeper than a MFP on average before complete reorientation.

From (\ref{eq:mean_z_drift}), we can compute $\langle z_\infty \rangle = \ell/(1-g)$, which is known as the ``transport'' MFP (tMFP) in the case of non-absorbing optical media.  
This quantity resurfaces and plays a crucial role in \S\ref{sec:VC_processes} below.

As carefully worded in the above, the answer to our question depends entirely on the scattering PF, and our theoretical arguments even bring that down to $g$.
Now, if asking about \emph{physical} rather than \emph{optical} depth, then the answer would also call for an effective extinction value, equivalently, an effective MFP, for the variable medium.

\subsection{Spatial resolution in the OS, from lateral dispersion based on sideways and forward scattering}

In App.~A, we investigate lateral dispersion of the scattered light field as well as the longitudinal drift examined in the above.
This enables us to quantify in 3D physical space the ``smearing'' effect discussed so far only as a directional phenomenon.

Returning to the RW-based formulation of RT in the OS, we computed exactly the mean position of the RWing particle along the original direction of propagation, namely, $\langle z_N \rangle$, after $N \ge 0$ scatterings.  
We must note however that the analytical estimate is strictly for a uniform \emph{unbounded} 3D space.
In fact, it is the possibility of large excursions of the RW into the $z < 0$ region that keeps $\langle z_N \rangle$ strongly bounded, and indeed makes it converge to a finite value $\langle z_\infty \rangle$ that we equated to the tMPF in the above.
To gain further insights, we also investigated in App.~A RWs in a uniform unbounded 3D \emph{half-}space, this time numerically: RWs are simply terminated if they enter $z < 0$ territory (see Fig.~A3 and corresponding discussion).  
The most drastic outcome is that, without the potentially very large negative-$z$ excursions, $\langle z_N \rangle$ now increases without bound with increasing $N$ (cf. Fig.~A3a).
That said, the analytical result we used in the above for $\langle z_N \rangle$ remains reasonably accurate up to $N \approx 6$ (hence $\langle \tau_N \rangle \approx 5$), which is roughly $n_\text{smear}$ when $g$ = 0.86.
This finding reinforces our prediction of the optical thickness of the OS based on $\langle \tau_{n_\text{smear}} \rangle$.

We can do the same \emph{mean} position estimation for horizontal directions, and indeed very easily: $\langle x_N \rangle = \langle y_N \rangle = 0$, after $N \ge 0$ scatterings, simply because of the rotational symmetry of the PF and its iterations by convolution in Fig.~\ref{fig:PF_convolve}.
Thus, to get to the interesting quantities, we now need to move on from 1$^\text{st}$-order statistics to their 2$^\text{nd}$-order counterparts, namely, variances $\langle z_N^2 \rangle-\langle z_N \rangle^2$ and $\langle x_N^2 \rangle$.
As seen in Figs.~Abc, we succeeded in getting exact results for the (full) unbounded 3D space only for $N$ = 0,1 and, for $\langle x_N^2 \rangle$, $N$ = 2.
The numerical investigation in App.~A shows that the 2$^\text{nd}$-order analytical model deteriorates much faster in $x$ than in $z$, in the sense of underestimation (as rationalized in App.~A).
The upgrade in the numerics from the full 3D space to the $z > 0$ half-space naturally makes no difference along the $x$- and $y$-axes.
However, by eliminating deviations into $z < 0$ space, it reduces the variance along the $z$-axis, which slightly favors the 2$^\text{nd}$-order analytical model.
It is noteworthy that the analytical models for variances require, like for the mean of $z$, knowledge of $g$, which comes from the 1$^\text{st}$-order coefficient in the Legendre expansion of the PF, and also for the 2$^\text{nd}$-order coefficient.

Quantitatively, variances in the position of the particles are nearly equal in all three axes for all but the first couple of steps when the full 3D space is accessible to their RWs.
The same is still true in the case of the half-space for as long as the mean $z_N$ is still in the radiative boundary layer (i.e., the analog of the OS for a uniform medium), specifically, for $N \lesssim n_\text{smear}$. 
For larger values of $N$, the half-space variance in vertical direction becomes smaller than that of its horizontal counterparts.
 
{\bf What does this tell us about visibility in the cloudy medium and for guidance in CT algorithm development?}
We are primarily interested here in the value of the standard deviation $\sqrt{\langle x_N^2 \rangle}$ while $\langle z_N \rangle$ is still in the radiative boundary layer, i.e., $\lesssim 5\ell$, hence $N \lesssim n_\text{smear}$ when $g \approx 0.86$.
That is a metric of lateral dispersion of a light beam that is not only collimated but also localized.
It therefore informs us about the spatial resolution that can be achieved when imaging through the cloudy optical medium.
To see this, visualize a laser source at the upper boundary of the half-space pointing vertically into the medium.
The resulting radiance field is, for all practical purposes, the Green function for the 3D RT problem, having a unitary Dirac-delta source in both physical and directional sub-spaces.

Inside the half-space, this laser light is broken into two components: ({\it i}) a directly-transmitted contribution that is still a $\delta$-function in the original direction, with a weight $\exp(-z/\ell)$ that decays rapidly with penetration depth $z$; ({\it ii}) a once- and more-scattered light field that can be identified as the point-spread function (PSF) of the cloudy medium, with a complementary weight that rapidly approaches unity when $z$ much exceeds the MFP $\ell$.
Both components are key to questions about imaging through the scattering medium.

Indeed, the imaginary laser source can be viewed as emitting ``reciprocal'' light propagating in reverse direction away from a given pixel at the sensor.
The direct component captures what the sensor can see in the absence of scattering, although reduced in strength by Beer's brutal law of exponential extinction.
The diffuse PSF component quantifies the ``importance'' for that pixel of all of the volume not along the direct beam, which quickly becomes the dominant source of light at the pixel of interest as $\tau = z/\ell$ increases well past unity.
In other words, at optical depths where the PSF overwhelms the direct light, spatial resolution is inherently limited to structures that are larger than the root-mean-square (RMS) of the lateral transport (in MFPs) via RW at optical depth $\tau$, namely, $\sqrt{\langle x^2 \rangle(\tau)}/\ell$.
Unfortunately, we do not have a closed-form expression for this important quantity.

However, the computational experiment resulting in Fig.~A3 enables us to make a statement about anticipated spatial resolution for CT in the OS.
The blue dots in the three panels indicate respectively: (a) $N \approx n_\text{smear} \approx 6$ at which $\langle \tau_N \rangle$ reaches 5, the predicted optical thickness of the OS; (b) the standard deviation of $\tau_N = z_n/\ell$ for the same $N$, which is $\approx$2.5; (c) same as (b) but for $x_N/\ell$, which is also $\approx$2.5.
Doubling that number yields an estimate of the full width of the PSF at the OS/VC interface.
That in turn is an indication in optical units of the size of a fluctuation in the extinction field that can be retrieved robustly by tomography at the edge of the VC, assuming that a large sensor pixel size is not a prior limitation.
At shallower optical depths, we anticipate that smaller structures can be reconstructed accurately.
At the edge of the cloudy medium, i.e., $\lesssim$1 optical depth, directly transmitted light is by definition a significant part of the signal.
Consequently, tomographic reconstruction methods will work the best in these optically shallow regions, and the spatial resolution of the 3D cloud imaging is determined only by the 2D sensor pixel scale.

In this last case of optically shallow structures in the extinction field, they are naturally the ones that drive the sharpest gradients in the cloud images (see Fig.~\ref{fig:emerging_features}).
In turn, those are precisely the ones picked up by feature-matching algorithms, and then used in CTH determination via stereo.

\subsection{Experimental validation}

Bohren et al. (\citeyear{Bohren_etal95}) ask: {\bf At what optical thickness does a cloud completely obscure the sun?}
We will argue that this question is closely related to the one about the optical depth of the effective OS/VC interface, to which we answer: $\approx$5, both empirically \citep{Forster_etal21} and theoretically (here).

Bohren and coworkers address their question experimentally using human subjects looking at a surrogate sun through a homogeneous artificial cloud.
Their ``cloud'' was composed of neutrally-buoyant spheres made of polystyrene (refractive index $\approx$1.59) with various size distributions (radii 0.652$\pm$0.0048~$\mu$m, 5.3$\pm$1.2~$\mu$m, 15.9$\pm$2.9~$\mu$m) suspended in a 0.26$\times$0.26$\times$0.50~m$^3$ tank filled with distilled water.
Mie computations (for the relative refractive index of 1.2) yield $g$-values of roughly 0.8, 0.85, and 0.9, respectively, for these PSDs.
Cloud optical thickness (COT) was gradually increased from 0 to the point where all three human observers agreed that they could no longer determine where the ``sun'' was located based on the transmitted light.

They did this by adding known numbers of spheres, and throughly mixing the contents of the tank.
They carefully measured COT (using Beer's law) at the 700~nm wavelength they used while the particle concentrations were low, and extrapolated to COT for higher ones using the calibrated linear relation.
The rule-of-thumb established by the authors was that a cloud with COT $\approx$ 10 will completely obscure the solar disk.
They found a weak trend toward larger COTs with larger particles (hence increased $g$), due to the enhanced forward scattering, but not enough to change the rule-of-thumb.

Bohren et al.'s ``solar disk vanishes at COT $\approx$ 10'' rule-of-thumb is consistent with our determination of an optical depth of $\approx$5 from the solar source forward and from the sensors backward to the OS/VC interface.
Indeed, for any identifiable structure in the cloud's extinction field to have a measurable (i.e., $>$noise) impact on remotely observable radiances it needs to be illuminated with at least partially collimated incoming sunlight and, similarly, be viewed with reflected light with at least residual directionality.

We determined in the above that, by examining iterated self-convolutions of the PF in both directions, the measured light cannot have suffered more than $\approx n_\text{smear}$ scatterings lest its directionality be seriously degraded.
That is precisely what lead to the prediction of $\approx$5 as the optical thickness of the OS (equivalently, optical depth of the VC).
It is therefore not a surprise that \emph{all} memory of directionality is lost (the solar disk is no longer detectable) at precisely twice that optical distance, namely, 10.
In short, we regard this alignment with Bohren et al. (\citeyear{Bohren_etal95}) as validation of our predictions based a controlled laboratory experiment.

\subsection{Partition of reflected sunlight between the OS and the VC}

Not all of the incoming sunlight reaches the OS/VC interface.
Some fraction is reflected back to space before that happens, and it can be estimated roughly, on the conservative side, by using the expression for transmittance from the $\delta$-Eddington version of the two-stream model \citep[e.g.,][]{MeadorWeaver80} in the absence of absorption:
\begin{equation}
\label{eq:T_deltaEdd}
T_g(\tau,\mu_0) = \frac{(2+3\mu_0)+(2-3\mu_0)\me^{-\tau/\mu_0}}{4+3(1-g)\tau},
\end{equation}
where $\mu_0$ is the cosine of the angle between the solar direction and the local outgoing normal to the plane-parallel cloud's outer boundary.  
Now, a real cloud's outer boundary is a flimsy fractal entity where droplets are condensing and evaporating all the while being flung around by turbulent motions over a wide range of scales.
Such a fractal surface has no well-defined normal.
However, one can imagine a ``running mean'' surface that cuts through the most tenuous outer layers of the cloud.
Alternatively, one can define the outer surface as the convex hull of the fractal boundary, allowing for regions of null extinction to exist inside the convex optical medium.

At any rate, we're interested in the value of $T_g(\tau,\mu_0)$ when $g = 0.85$ and $\tau = 5$.
It ranges from $\approx$0.32 to $\approx$0.82 for $0 < \mu_0 \le 1$, recalling that the local irradiance is $\propto\mu_0$.
So, to get our rough estimate of how much sunlight ever reaches the VC, we are interested in an average of $T_{0.85}(5,\mu_0)$ over the illuminated portion of the cloud's surface.
If the distribution of $\mu_0$ across the surface is uniform on (0,1], then the mean $T$ is $\approx$0.57.
The proverbial spherical cloud leads to $\approx$0.64.
Our rule-of-thumb can be $\approx$60\%.
Factoring in the spatial variability of the OS, which is known to lead to an underestimate of transmittance \citep[e.g.,][]{cahalan94}, we can settle on $\approx$2/3.
{\bf Consequently, we reckon that $\approx$1/3 of the incoming sunlight never enters the VC.}
To a first approximation, that is the light that forms the ``features'' in the cloud imagery on (or near) the illuminated side of the cloud.
Let us further examine the fate of the light that does \emph{not} enter the VC by revisiting Fig.~\ref{fig:Koch_in+out_VC}.

Figure~\ref{fig:Koch_in+out_VC} builds on results from \cite{Forster_etal21} to visualize the VC and OS in a toy cloud model based on a fractal outer shape and a scale-invariant stochastic model for the inner structure that mimics turbulence overlaid with cloud-scale vertical gradient that captures convectively-driven stratification of the extinction field.  
COT down the central column is 40. 
Version (a) shows the solar rays penetrating the cloud from the West ($\theta_0$ = 60$^\circ$).
They are color-coded by optical distance $\tau_0$ to the solar source, and terminated at $\tau_0$ = 5.
Version (b) shows the OS in color and the VC in grey, as defined by the nine MISR view angles, although the three dominant cameras are Df, An and Da.

We see that there are lots of locations where the optical distance to the sun and to the cameras are both relatively small.
This is an opportunity for singly-scattered light to reach the sensor and convey the sharpest possible imprint of physical cloud structure as an image feature.
We note however that there is not much incoming/outgoing overlap for the most oblique forward-looking camera (Df) and, due to this particular cloud and illumination geometry, it occurs only very near the top of the cloud.
There is increasing overlap for the less oblique forward-looking cameras, and maximal overlap for all the aft-looking cameras.
The strong forward peak of the PF allows for still quite sharp imaging of structures in somewhat deeper layers thanks to light that has suffered as much as a handful of scatterings.
However, the more scatterings suffered, the less structured the light field.
It will therefore convey less of the information about cloud structure that is exploited in the inverse imaging problem in CT.

\section{Diffusional transport across the VC}
\label{sec:VC_processes}

\begin{figure}[htbp]
\begin{center}
\includegraphics[width=3.14in]{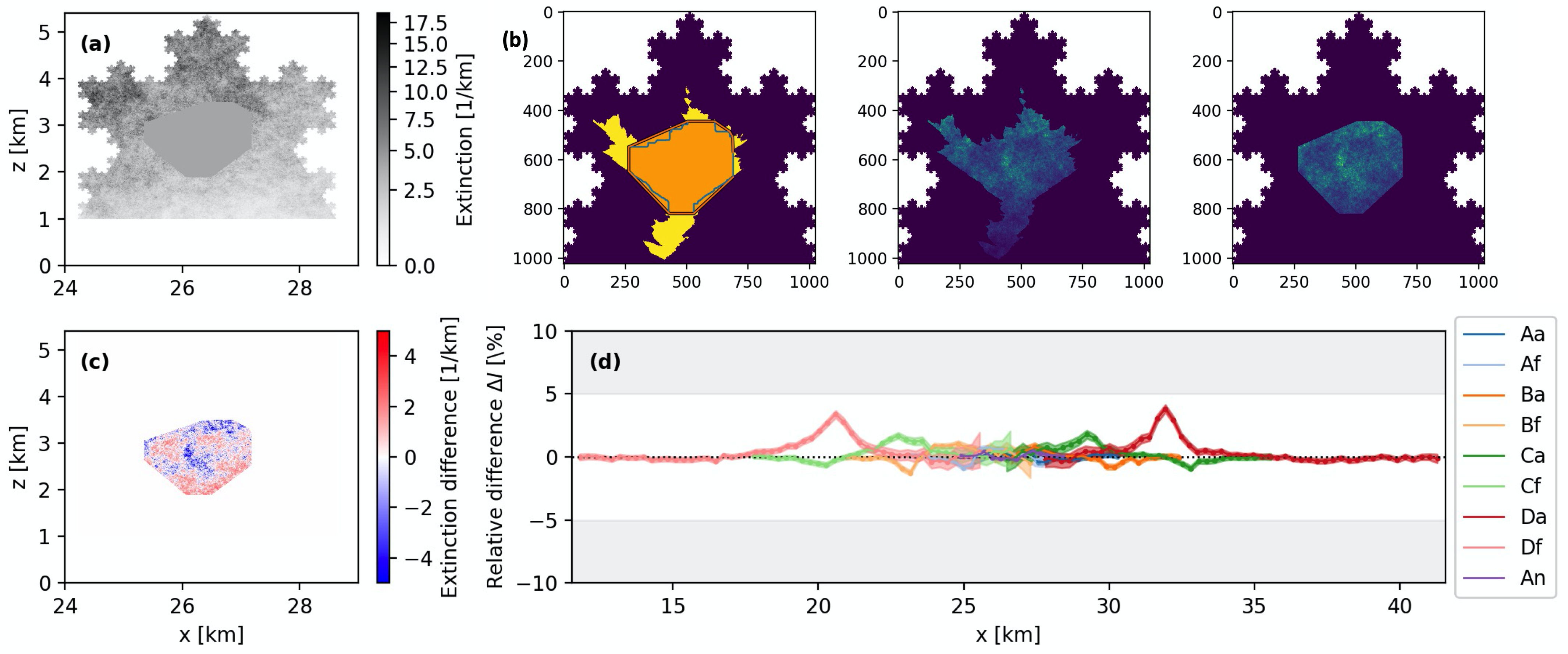}
\caption{
(a) The Koch \emph{fractal} cloud by \cite{Forster_etal21} in Fig.~\ref{fig:Koch_in+out_VC}, but with optical thickness 20 (down from 40) along the central column, and its \emph{adaptive} VC identified as the \emph{regular} internal region; the uniform extinction value therein is the mean taken inside the VC.
(b) Genesis of the adaptive VC, from left to right: 
a VC mask in yellow, based on the threshold optical depth of 4 (down from the usual 5) and, in orange, a convex set circumscribed by the yellow VC mask after a couple of ``erosion/dilation'' operations \citep[e.g.,][]{JackwayDeriche96} that boost the \emph{mean} optical depth of the VC back to 5-or-so; 
internal turbulence field inside the fractal yellow VC; 
the same, but for the convex orange VC.
(c) Same internal turbulence field inside the convex orange VC, with mean from panel (a) removed.
(d) MYSTIC-based high-precision estimate of radiance differences between original Koch cloud and its counterpart with turbulent extinction field replaced by its mean value inside the convex orange VC; phase function for $r_\re = 10 \mu$m, SZA = 0$^\circ$.
}
\label{fig:adaptive_VC}
\end{center}
\end{figure}

So far, we have defined and visualized the VC as a complex domain inside the Koch cloud that inherits a fractal structure from it due to the solar and reciprocal sensor beams drilling into the medium, as illustrated in Fig.~\ref{fig:Koch_in+out_VC}.
In sharp contrast, \cite{Forster_etal21} performed their numerical experiments using a simple geometric shape to approximate the VC inside the fractal Koch cloud.
Figure~\ref{fig:adaptive_VC} shows an adaptive compromise between these two operative definitions.
Although the VC definition threshold is lowered from 5 to 4, the resulting VC is then shrunk into a \emph{convex} shape circumscribed by a smoothed version of the original fractal VC.
This brings the optical distance from the solar source and all 9 MISR cameras to the adaptive VC boundary back to $\approx$5.
The procedure is illustrated in Figs.~\ref{fig:adaptive_VC}a--c.
Figure~\ref{fig:adaptive_VC}d shows that the radiance field differences between the original cloud and the simplified version in Fig.~\ref{fig:adaptive_VC}a are inside the 5\% tolerance.

Imposition of a convex shape onto the VC is rationalized further on, based on computational considerations.
By the same token, we see that the yellow version of the VC, along with some of the regularly-shaped VCs used by \cite{Forster_etal21}, reaches the very bottom of the cloud, by its definition based on the solar source and overhead sensors.
The final orange version, however, does not touch cloud base.
That is also a desirable property of the VC for physical and computational reasons since, there too, the RT will be influenced by the presence of the lower boundary.
3D RT codes need to account correctly for the gradual transition from multiple scattering to pure streaming, even if the light is diffuse rather than collimated.
However, there are neither strongly-directional sources nor well-collimated sensors that see the cloud base in the present context of passive space-based cloud remote sensing.
Consequently, the required optical thickness of the OS between cloud base and VC will probably not need to be 5-to-6, but some lower number to be determined in a future study.

\subsection{Angular distribution of radiance in the VC and associated RW in its 3D volume}
\label{sec:DA_from_RT}

Looking back at Fig.~\ref{fig:PF_convolve}, we clearly see that the sunlight that has filtered through the OS, i.e., predominantly with $N \gtrsim n_\text{smear}$ ($\approx$6.5 for $g$ = 0.86), and is impinging on the VC has been highly smoothed angularly after multiple scatterings. 
This radiance distribution in direction space can be represented by ever more truncated Legendre series.  
For single scattering (gray curve), we summed 627 terms while, after the 10$^\text{th}$ scattering (red curve), the iterated phase function is already approximated reasonably well with the first two terms.
Formally, we are stating that $p(\theta)^{(*10)} \approx 1+3g_{10}\cos\theta$, where $g_{10} = g^{10} = 0.22$ and the superscript ``$*(10)$'' means ``10 times self-convolved.''

This suggests that (optically) far enough from collimated light sources and other regions where we anticipate the light will be streaming ballistically (near boundaries), we can approximate the radiance field $I(\bx,\bOmega)$ with just two spherical harmonics:
\begin{equation}
\label{eq:I_diffusion}
I(\bx,\bOmega) \approx \frac{1}{4\pi} \left[ J(\bx) + 3\bOmega\cdot\bF(\bx) \right],
\end{equation}
where $\bx$ is position in the optical medium M, $\bOmega$ is direction of propagation, and
\begin{eqnarray}
\label{eq:J_scalarF}
J(\bx) &=& \int_{4\pi} I(\bx,\bOmega) \, \dif\bOmega \\
\label{eq:F_vectorF}
\bF(\bx) &=& \int_{4\pi} \bOmega \, I(\bx,\bOmega) \, \dif\bOmega 
\end{eqnarray}
are respectively the scalar and vector radiative flux fields.  
Similarly, we use a 2-term harmonic expansion of the PF:
\begin{equation}
\label{eq:p_diffusion}
p(\bOmega\cdot\bOmega^\prime) \approx \frac{1}{4\pi} \left[ 1 + 3g \bOmega\cdot\bOmega^\prime \right].
\end{equation}
These natural assumptions have far-reaching ramifications for the 3D RT in sufficiently opaque clouds to justify them.

The general steady-state 3D RT equation in a uniform but arbitrarily shaped scattering medium M is \citep{Chandra1950,mishchenko2002vector}
\begin{equation}
\label{eq:3D_RTE}
\left[ \bOmega\cdot\bnabla + \sigma_\re \right] I(\bx,\bOmega) = 
\sigma_\rs \int_{4\pi} p(\bOmega\cdot\bOmega^\prime) I(\bx,\bOmega^\prime) \frac{\dif\bOmega^\prime}{4\pi},
\end{equation}
where $\sigma_\re$ and $\sigma_\rs \, (\le \sigma_\re)$ are respectively extinction and scattering coefficients.
This integro-differential RT equation (RTE) balances sinks (left-hand side, advection and extinction) and sources (right-hand side, in-scattering) of light in a small volume around $\bx$ in 3D space propagating into a small cone centered on direction $\bOmega$ on the 2D unit sphere.
For the present application, we assume that all light sources are isotropic and confined to the boundary of M, denoted ``$\partial$M:''
\begin{equation}
\label{eq:iso_BCs}
I(\bx,\bOmega) = f(\bx) |\bOmega\cdot\bn(\bx)|/\pi,\text{ for }\bx\in\partial\rM\text{ and }\bOmega\cdot\bn(\bx)<0
\end{equation}
where $\bn(\bx)$ is the outgoing normal of $\partial$M at point $\bx$, and $f(\bx)$ is the given incoming irradiance.

Taking the integrals $\int_{4\pi}[\cdots]\dif\bOmega$ and $\int_{4\pi}\bOmega[\cdots]\dif\bOmega$ of the RTE in (\ref{eq:3D_RTE}), using (\ref{eq:I_diffusion})--(\ref{eq:p_diffusion}) and the PF normalization, we obtain coupled 1$^\text{st}$-order partial differential equations (PDEs):\footnote{
There are other paths from the RT equation in (\ref{eq:3D_RTE}), potentially with spatially variable optical properties $(\sigma_\re,\sigma_\rs)$ and $p(\cdot)$, to the diffusion model in (\ref{eq:J_scalarF})--(\ref{eq:F_vectorF}). 
The most illuminating one is probably via asymptotic analysis of the RT equation in the limit of scaling $(\sigma_\re/\epsilon,\sigma_\rs/\epsilon)$ when $\epsilon \to 0$ \citep{Larsen1980,Pomraning1989}.
}
\begin{eqnarray}
\label{eq:conservation}
\bnabla\cdot\bF(\bx) &=& -\sigma_\ra J(\bx) \\
\label{eq:constitutive}
\bF(\bx) &=& -\frac{\ell_\rt}{3} \bnabla J(\bx)
\end{eqnarray}
where $\sigma_\ra = \sigma_\re-\sigma_\rs \in [0,\sigma_\re]$ is the absorption coefficient, and $\ell_\rt$ is the \emph{transport} MFP (tMPF) introduced in App.~A using RWs on the 2D sphere in the absence of absorption ($\sigma_\ra = 0$).
Here, we are including the potential for absorption, and find
\begin{equation}
\label{eq:MFP_t}
\ell_\rt = \frac{1}{\sigma_\rt} = \frac{1/\sigma_\re}{1-\omega g} = \frac{\ell}{1-\omega g},
\end{equation}
where $\sigma_\rt = (1-\omega g)\sigma_\re$, is the \emph{transport} or ``scaled'' extinction coefficient, and $\omega = \sigma_\rs/\sigma \in [0,1]$ is the single scattering albedo (SSA).
In (\ref{eq:conservation}), we have the \emph{exact} expression of radiance energy conservation.
In (\ref{eq:constitutive}), we have the constitutive law (a.k.a. Fick's law) that defines the radiative diffusion \emph{approximation}.
Thus, in (\ref{eq:constitutive}), we identify $D = \ell_\rt/3$ as the (steady-state) diffusivity coefficient.

Finally, based on (\ref{eq:iso_BCs}), the BCs for the diffusion transport model in (\ref{eq:conservation})--(\ref{eq:constitutive}) is completed with
\begin{equation}
\label{eq:BCs_DA}
\frac{1}{4}\left[ J(\bx) - 3\chi\bn(\bx)\cdot\bF(\bx) \right] = f(\bx), 
\end{equation}
where $\chi$ is the so-called ``extrapolation length'' factor. 
The common (Marshak) assumption\footnote{
We assume for simplicity a purely absorbing BC in (\ref{eq:iso_BCs} and \ref{eq:BCs_DA}), i.e., $f(\bx) = 0$.  
Without loss of generality, we can take $\bn$ along the $z$-axis at any given boundary point. 
Then the incoming/outgoing radiant energy fluxes at said boundary point are, based on (\ref{eq:I_diffusion}), $J/4 \mp F/2$, where $F = \|\bF\|$. 
Thus, in the case of no incoming flux, $J/F = 2$. 
Substituting Fick's law (\ref{eq:constitutive}), we find $J/|\nabla J| = (2/3)\ell_\rt$. 
By identifying with the definition of the extrapolation length factor $\chi$ at a boundary point, $J/|\nabla J| = \chi\ell_\rt$, we find $\chi$ = 2/3.
In practice, this tells us the distance from the boundary at which $J$ would vanish if it followed a linear trend.
This, in turn, tells how much extra volume we have to add to the medium M if we wanted to replace the exact Robin BCs in (\ref{eq:BCs_DA}) by Dirichlet counterparts, which is sometimes a convenient approximation.
Other arguments lead to other values for $\chi$; see \cite{CaseZweifel1967}.
}
is $\chi = 2/3$.
Equation (\ref{eq:BCs_DA}) expresses a 3$^\text{rd}$-type (a.k.a. Robin) BC that combines both density $J(\bx)$ and current $\bn(\bx)\cdot\bF(\bx)$ at the domain's boundary ($\bx\in\partial$M).
The (charge) ``density'' and (electrical) ``current'' terminology respectively for $J$ and for $\bF$ is widely used but clearly borrowed from electrodynamics where (\ref{eq:conservation}) expresses conservation of charge, and (\ref{eq:constitutive}) is Ohm's law.
Although we are advocating for particle diffusion and RWs, transport of charge through a conductor is a valid analog of light transport in the VC where the gradient in illumination between the sun-exposed and self-shading sides replaces the imposed drop in potential.
An even better analog is the transport of a liquid through a porous medium, where (\ref{eq:conservation}) expresses conservation of mass, and (\ref{eq:constitutive}) is D'Arcy's law, with pressure head playing the role of the illumination differential across the VC.
A good example in geophysics is ground water flow in the vadose zone \citep[e.g.,][]{groundwater}.

Substituting (\ref{eq:constitutive}) into (\ref{eq:conservation}), we get the standard 2$^\text{nd}$-order elliptical Helmholtz PDE:
\begin{equation}
\label{eq:Helmholtz_PDE}
\left[ \bnabla^2 + \frac{3\sigma_\ra}{\ell_\rt} \right] J = 0,
\end{equation}
subject to BCs $\left. \left[ 1 - \chi\ell_\rt\bn(\bx)\cdot\bnabla \right] J(\bx) \right|_{\bx\in\partial\rM} = 4f(\bx)$, from (\ref{eq:BCs_DA}) and (\ref{eq:constitutive}).
In the absence of absorption ($\sigma_\ra = 0$), that PDE reduces to the classic Laplace PDE, $\bnabla^2 J = 0$, which describes the steady-state 3D RW (a.k.a. Brownian motion) of particles from boundary sources (where $f(\bx) > 0$) to boundary sinks (anywhere, but especially on the non-illuminated side of the VC). 

The PDE in~(\ref{eq:Helmholtz_PDE}) defines an important length scale in radiation transport theory, namely, the diffusion length:
\begin{equation}
\label{eq:Diffusion_length}
L_\dif = \sqrt{\frac{\ell_\rt}{3\sigma_\ra}} =  \frac{\ell}{\left[ 3(1-\omega)(1-\omega g) \right]^{1/2}}. 
\end{equation}
$L_\dif$ is the characteristic scale of the exponential terms in the fundamental solution of (\ref{eq:Helmholtz_PDE}).
In essence, $L_\dif$ is the distance from (in this case, boundary) sources at which light is severely extinguished by way of the absorption by droplets that occurs at every (now non-elastic) scattering. 
Table~\ref{tab:MODIS_SWIR_chans} lists the optical properties in (\ref{eq:Diffusion_length}) across 3 MODIS SWIR channels, along with the extinction coefficient for a nominal LWC of 0.1~g/m$^3$ (assuming the same PSD as in Fig.~\ref{fig:PF_convolve}); MODIS's red channel is also displayed  for reference.
While neither the extinction (hence COT or MFP) nor $g$ vary much across the VNIR-SWIR spectral range, SSA $\omega$ does, in the sense that the single scattering co-albedo $(1-\omega)$ that appears in (\ref{eq:Diffusion_length}) varies by orders of magnitude.
Thus $L_\dif$ varies significantly across the MODIS SWIR channels, while $\ell$ does not.

\begin{table}[htp]
\caption{Wavelength, MODIS channel \#, extinction coefficient (assuming LWC = 0.1~g/m$^3$), SSA, asymmetry factor, and diffusion length in MFPs from (\ref{eq:Diffusion_length}) for the cloud droplet PSD used in Fig.~\ref{fig:PF_convolve}. First column is color-coded to reflect choices in Fig.~\ref{fig:Diffusion_length_Koch}.}
\small
\begin{center}
\begin{tabular}{|c|c|cccr|}
\hline
\hline
$\lambda$ & channel & $\sigma_\re$ & $\omega$ & $g$ & $L_\dif/\ell$ \\

  [nm]    & \#      & [1/km]       & [-]      & [-] & [-]           \\
\hline
\tb{645}  & 1 & 15.82 & 0.999996 & 0.8610 & 774.3 \\
\tg{1240} & 5 & 16.14 & 0.998042 & 0.8499 &  33.5 \\
\ty{1640} & 6 & 16.51 & 0.991238 & 0.8456 &  15.3 \\
\tr{2130} & 7 & 16.70 & 0.971810 & 0.8418 &   8.1 \\
\hline
\end{tabular}
\end{center}
\normalsize
\label{tab:MODIS_SWIR_chans}
\end{table}

\begin{figure}
\begin{center}
\includegraphics[width=3.14in]{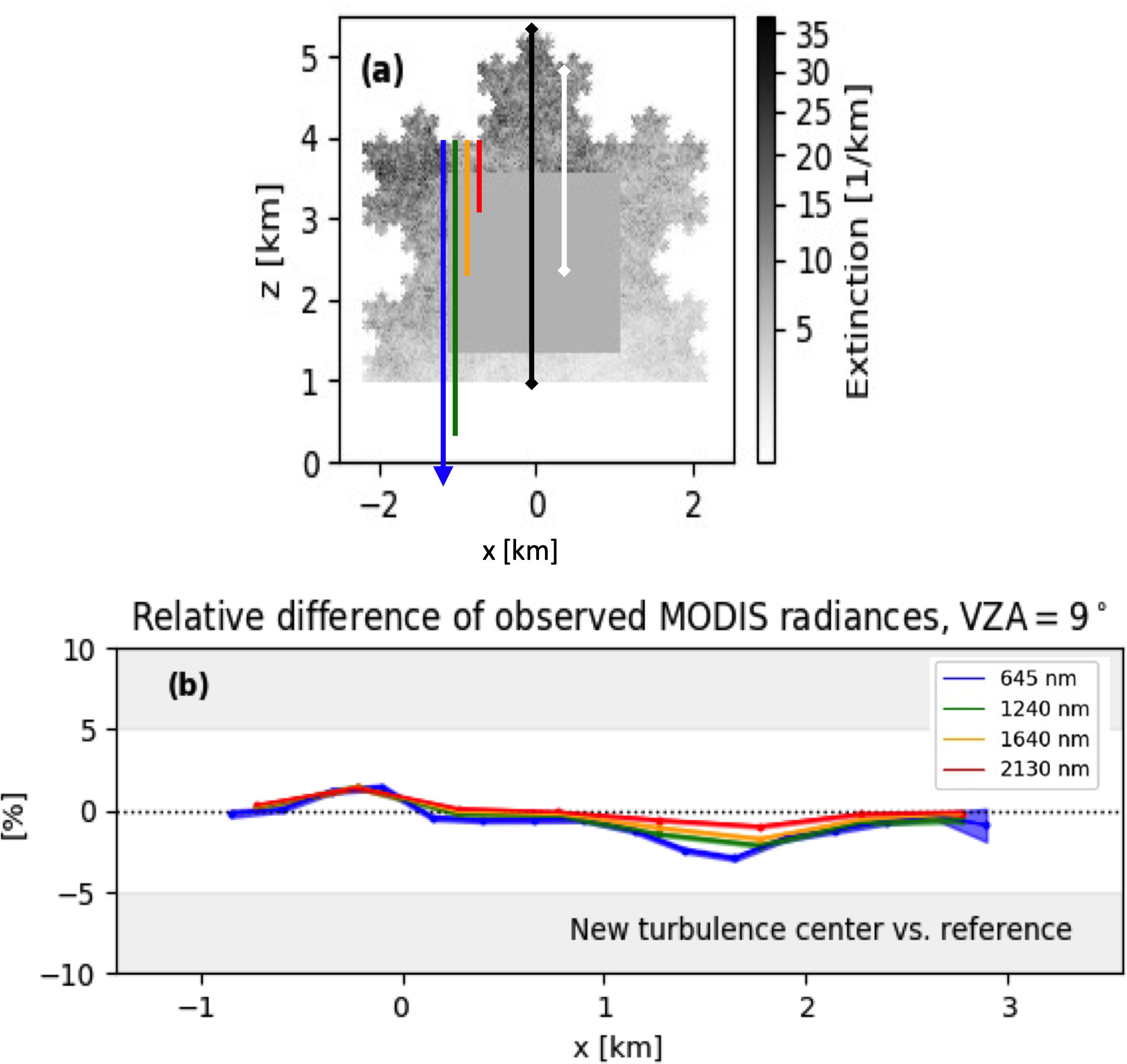}
\end{center}
\caption{
{\bf (a)} Various length scales are shown graphically for the Koch fractal cloud in Fig.~\ref{fig:Koch_in+out_VC} with COT = 40 along the central column (denoted $\tau_\text{central}$), which is highlighted with a 4-km long black diamond-ended line. 
For comparison, the horizontally-averaged COT is 22.4 and represented with the accordingly shorter white diamond-ended line. 
Taking $\ell$ as the inverse of the mean extinction in (\ref{eq:Diffusion_length}), we use 4 lines to show the extents of $L_\dif$ color-coded as in Table~\ref{tab:MODIS_SWIR_chans} and in panel (b).
{\bf (b)} Reproduced for context from \cite{Forster_etal21}, Fig.~9b, SZA = 0$^\circ$ and VZA = 9$^\circ$ (average for MODIS across the MISR swath). 
It shows the along-track relative differences between radiances observed by MODIS for two Koch clouds that differ only in the detailed structure of the extinction field inside the grey rectangle in panel (a).
Differences are well within the noise-determined threshold for significance in CT at 5\%.
The grey rectangle (and variations thereof) is used extensively by \cite{Forster_etal21} as a proxy for the VC.
}
\label{fig:Diffusion_length_Koch}
\end{figure}

Figure~\ref{fig:Diffusion_length_Koch} illustrates $L_\dif$ across MODIS SWIR wavelengths in the present context of VC and OS characterization, as a prelude to CT with MISR and MODIS.
We see that, while $L_\dif$ at the essentially non-absorbing wavelength of 645~nm is off the charts, its magnitude quickly shrinks as wavelength increases into the SWIR region.
Specifically, at 1240~nm, $L_\dif$ is seen in panel (a) to be commensurate with the overall size of the cloud, so we anticipate very little difference in the already low sensitivity to VC structure obtained at 645~nm; panel (b) confirms this prediction.
At 1640~nm, $L_\dif$ is roughly halved from 1240~nm, and as a result the light is strongly absorbed in the OS and as it enters the VC; sensitivity to VC structure takes a major hit, as is seen in panel (b).
Finally, at 2130~nm, $L_\dif$ is just a little more than the typical thickness of the OS (hence depth of the VC); sensitivity to a change in the VC is all but gone because the VC will be essentially dark due to the intense absorption in the outer layers.
The Koch cloud used here has a COT of 22.4 on average along the $x$ dimension.
Optically thinner clouds won't have much of a VC.
In optically thicker clouds, $L_\dif$ values across the SWIR channels are reduced in proportion, while the VC is increasing in volume.
All the sensitivities to VC details are therefore further diminished.

In summary, we can think of the VC as a region M inside the cloud where ``particles'' representing units of radiant energy (at a non-absorbing wavelength) wander randomly in the optical medium M from an entry point on $\partial$M to an escape point on $\partial$M.
From that standpoint, the VC can also be called the ``diffusion zone'' in an opaque 3D cloud where the light arrives after enough scatterings to smooth radiance directionality down to an isotropic term and a dipole term in (\ref{eq:I_diffusion}), and then scatters many more times before emerging.

The RW in the VC is characterized by effectively isotropic scatterings and steps of mean length $\ell_\rt$, which is $(1-g)^{-1}$ times longer then the usual MFP $\ell = 1/\sigma_\re$, which in turn is the mean of the exponential distribution of steps (a.k.a. free paths) associated with Beer's law of extinction.
Consequently, each \emph{effective} isotropic scattering counts, on average, for $(1-g)^{-1}$ scatterings according to the forward-peaked PF.
Recall that, for $g$ = 0.86 (from Fig.~\ref{fig:PF_convolve}), we have $(1-g)^{-1}$ = 7.14 as scaling factor for the RW, in both the size of the steps across 3D space and the (discrete) time it takes.
In the presence of absorption by cloud particles, the SSA $\omega$ multiplies $g$ in the above forward-peaked PF considerations: $(1-g)^{-1}$ becomes $(1-\omega g)^{-1}$, which changes little since $\omega$ is only slightly less than unity in the SWIR region.
However, in the full-blown diffusion regime the effect of absorption is highly amplified in the highly scattered radiation.
Light is quickly extinguished at distances from sources (in our case, the boundary of the VC that faces the Sun) that exceed the ``diffusion length'' $L_\dif$, which scales with $(1-\omega)^{-1/2}$ in (\ref{eq:Diffusion_length}).

\subsection{Control of cloud-scale contrast in observable radiances by VC}

To understand the primary imprint of the VC on cloud images, we use in App.~B a tutorial cloud model that is \emph{all} VC.
In other words, the transport regime is diffusive in the entire cloud, a reasonable approximation for clouds with very large ($\gg$5) optical thickness in all three dimensions.

To explore remote sensing of broken clouds from potentially very oblique view angles, \cite{davis2002} sought a 3D cloud geometry that is not plane-parallel yet remains analytically tractable, at least in the diffusion limit, which he found in the perfect sphere.
The author adopted the natural definitions of reflectivity $R$ and transmissivity $T$ as outgoing fluxes integrated respectively over the illuminated and shaded hemispheres, as a special case of the general formulation by \cite{DavisKnyazikhin2005}, and normalized by the solar flux through the area of the cloud projected perpendicular to the incoming beam.
\cite{davis2002} showed that, in the absence of absorption,
\begin{equation}
\label{eq:RoT_ratio}
\frac{R}{T} = \frac{(1-g)\tau}{2\chi},
\end{equation}
where $\tau$ is the optical diameter of the spherical cloud.
This diffusion-based prediction was validated with Monte Carlo simulations of RT in spherical media with varying $\tau$.
Indeed, as $\tau_\rt = (1-g)\tau$ reaches and exceeds unity, the predicted linear increase of $R/T$ with $\tau$ proves accurate when $\chi$ was set to 2/3.
The \cite{HenyeyGreenstein41} PF was adopted, with $g = 0.85$.
The uniform collimated solar beam impinges on the sunny side of the cloud boundary at an angle varying between 0 and $\pi$/2 depending on the angular distance $\vartheta$ from the sub-solar point, as viewed from the center of the sphere.
Two source models were then used: the light either continues into the cloud without change in direction, or it is sent into the cloud at a random direction, as in (\ref{eq:iso_BCs}) with $f(\bx) \propto \cos\vartheta(\bx)$.
As $\tau$ increases, the diffusion-theoretical prediction for $R/T$ in (\ref{eq:RoT_ratio}) is approached at similar rates for both boundary sources, but from opposite directions.

\cite{davis2002} notes that, if the cloud is an infinite plane-parallel slab, then the $R/T$ contrast ratio is again as in (\ref{eq:RoT_ratio}) where $\tau$ is then the optical thickness of the cloud.
Spheres and slabs are, in essence, ellipsoids with respectively three identical finite semi-axis and one finite with two infinite semi-axis.
This leads to the prediction that cylinders (ellipsoids with two identical finite semi-axis and one infinite one) should have the same $R/T$ contrast ratio.
This is prediction is confirmed in App.~B using, for simplicity, circular media in 2D RT and 2D diffusion theory.
Therein, 2D Monte Carlo simulations are used to verify that the accuracy of associated 2D diffusion-theoretical prediction becomes reasonable when $(1-g)\tau$ becomes {\it O}(1) and continues to improve as $\tau$ continues to increase (cf. Fig.~B2).

\begin{figure}
\begin{center}
\includegraphics[width=3.14in]{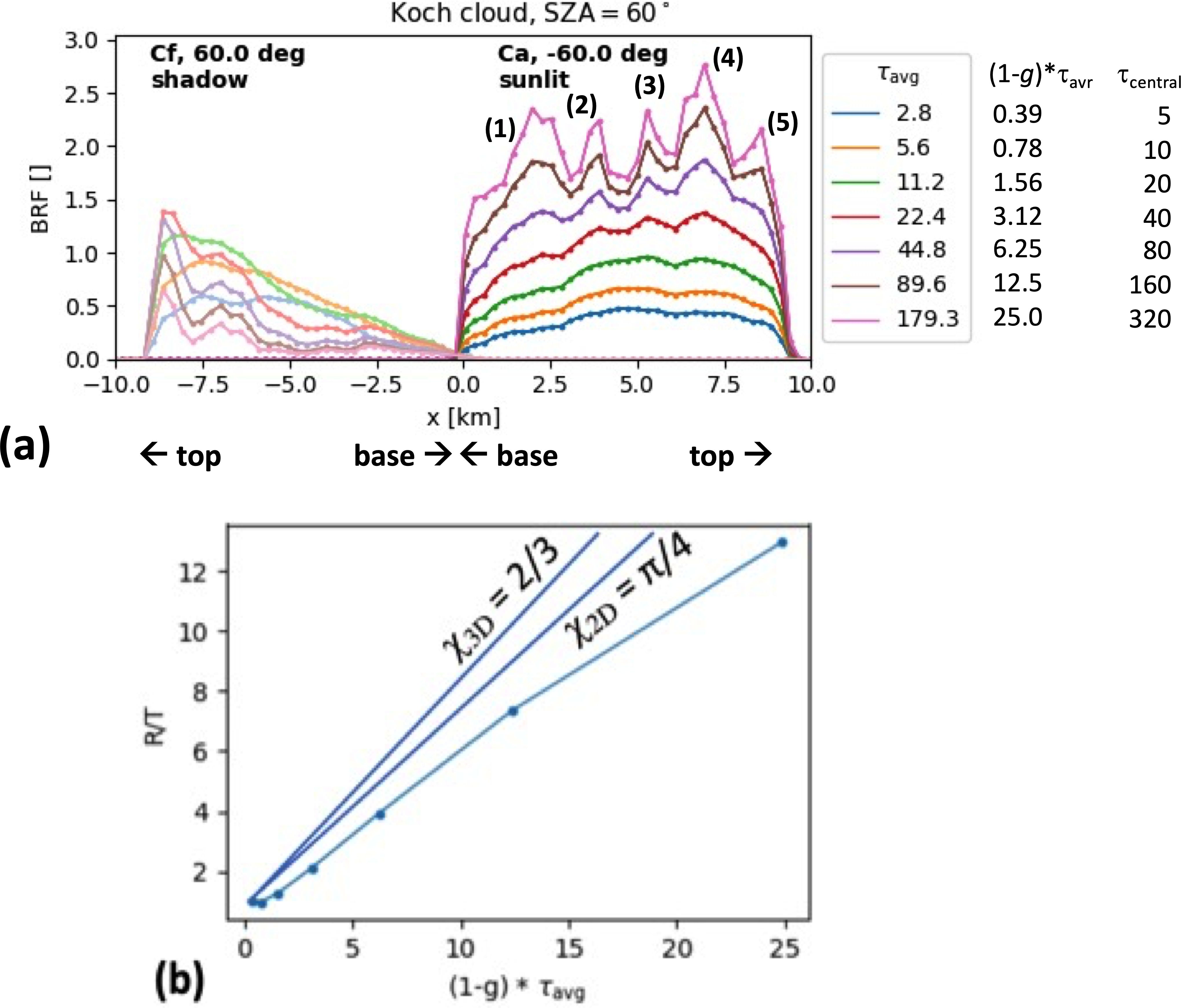}
\end{center}
\caption{
{\bf (a)} The same 2D Koch fractal cloud model as in Fig.~\ref{fig:Koch_in+out_VC} is used with an opacity that is doubled 5 times, starting at $\tau_\text{central}$ (integrated extinction down the middle column) = 5.  
To the right, $\tau_\text{central}$ is translated into $\tau_\text{avg}$ (horizontally averaged optical thickness), and its scaled value $(1-g)\tau_\text{avg}$, where $g = 0.86$.
Numbered image ``features'' will be discussed further on (\S\ref{sec:concl}).
{\bf (b)} Rough estimates of $R$ and $T$ are obtained by averaging the radiances (expressed in non-dimensional Bidirectional Reflection Function or ``BRF'' units) across the cloudy pixels, respectively in the Ca and Cf cameras.
Thus a proxy for the $R/T$ contrast ratio is computed and plotted against $(1-g)\tau_\text{avg}$ yielding a near-linear relation, as predicted for clouds with simple shapes (slabs, spheres, cylinders) that are all VC.
Predicted slopes for 3D and 2D diffusion models are indicated for reference, but not expected to apply directly to this $R/T$ metric, as explained in the main text.
}
\label{fig:Koch_R_vs_T}
\end{figure}

How does the sunny-to-shady-side contrast ratio in (\ref{eq:RoT_ratio}) fare with more realistic clouds in terms of both outer and inner structure?
Figure~\ref{fig:Koch_R_vs_T} uses the 2D Koch fractal cloud used by \cite{Forster_etal21} with SZA = 60$^\circ$, coming from the West, and VZA = $\pm 60^\circ$, i.e., the equivalents of MISR's Cf and Ca cameras.
In this configuration, the Cf camera sees only the shaded side of the cloud, and the cloud image is registered to ground pixels West of the cloud.
By the same token, the Ca camera sees all of the illuminated side of the cloud in exact retro-scattering geometry, and the cloud image is registered to ground pixels East of the cloud.
Although we are dealing with radiances rather than fluxes, we can still integrate them across the two cloud images in Fig.~\ref{fig:Koch_R_vs_T}a: Cf yields a proxy for $T$ while Ca yields one for $R$.
Figure~\ref{fig:Koch_R_vs_T}b displays the proxy $R/T$ ratio, and we can see that it is a monotonic function of $\tau$ (averaged horizontally) that is not far from a linear trend as soon as the cloud is optically thick enough to develop a VC ($\tau_\text{central} \gtrsim 10$).
$R/T$ is off-trend for the most opaque incarnation ($\tau_\text{central} = 320$), but that may be due to the Monte Carlo noise in the estimation of $T$.
The slope is shallower than for the diffusion theory predictions for circular or spherical scattering media.
However, there is no reason to expect them to agree since, apart from the fractal outer shape here, the theory is for angularly integrated radiance while the simulation yields a single sampled viewing direction each for the $R$ and $T$ proxies.
Moreover, the analytical predictions are for strictly uniform clouds.
We know that, for a fixed optical thickness, heterogeneous clouds like this one yield lower $R$ and higher $T$, hence lower $R/T$, as observed here.

Radiances from real 3D clouds with complex outer geometries cannot be so easily partitioned into ``$R$'' and ``$T$'' pixels.
Looking ahead, such clouds observed by present and future space-based multi-angle sensors will have \emph{effective} ``$R$'' and ``$T$'' values obtained from the appropriate image segmentation algorithms tuned to separate sunny and shady cloud boundaries at the sensor's pixel scale.
From there, an effective cloud-scale $R/T$ ratio can be formed. 
Realistic LES-generated clouds can be used to train the segmentation algorithm and calibrate the monotonic relation between observed $R/T$ and the mean $\tau$.
From the CT perspective, this yields an initial estimate of mean cloud extinction since the cloud's volume will have been constrained, for instance, by ``space-carving,'' (i.e., the volumetric intersection of the 2D cloud masks across all observing angles, cf. \cite{Lee_etal18}).

\subsection{Lateral radiation transport along the OS/VC interface}

We have established that a surplus of \emph{diffuse} light impinges on the part of the VC boundary that is exposed to the Sun, and that there is a deficit on the opposite side of the VC facing the self-shaded part of the cloud.
This cloud-scale gradient in radiant energy density drives an overall flow through the cloud's VC and, for that matter, the whole cloud.
That said, for the purposes of physics-informed CT algorithm design, we need to characterize the lateral transport of radiant energy along the boundary of the VC since that will define the spatial smoothing of the radiance field by the VC itself.
To answer this question quantitatively, we will mine the literature on cloud PSFs in the diffusion regime.
Interestingly, there is a bifurcation between research interested in reflected and transmitted radiation due to the different remote sensing technologies of interest.

On the reflection/back-scattering side of the cloud, the technology driver was MUliple-Scattering Cloud Lidar (MUSCL), an active cloud probing methodology that was being explored in the early 2000s.
Before that, multiple scattering lidar experiments \citep{Flesia1995MUSCLE} were primarily about capturing the effects of one or more forward-scattering events not accounted for in the standard lidar equation but present in lidar returns due to the finite field-of-view (FOV) of the receiver.
This gradual deviation from direct propagation not unlike our OS concern in App.~A and \S\ref{sec:OS_processes}.
Breaking away from that trend, Davis et al. (\citeyear{davis1999,davies1999}) developed the diffusion-theoretical framework for predicting MUSCL signals and their information content in terms of cloud properties.
A quantity of particular interest to the present investigation is the RMS radius of the reflected Green function, equivalently, the RMS distance between entry and escape positions of the RWing ``particle'' carrying units of radiant energy in the diffusion model, which is tantamount to an estimation of the cloud PSF half-width.
Let $\rho_{\text{rms},R}$ be the RMS entry-to-escape distance along the illuminated boundary.
For sufficiently opaque media, diffusion theory predicts that, in the absence of absorption, we have:
\begin{equation}
\label{eq:RMSrho_R}
\rho_{\text{rms},R} \propto \sqrt{\ell_\rt H} = H/\sqrt{\tau_\rt},
\end{equation}
where $H$ is the physical thickness of the medium.
The exact prefactor and pre-asymptotic corrections in $\ell_\rt/H = 1/\tau_\rt$ for (\ref{eq:RMSrho_R}) were derived in closed-form for plane-parallel slabs, and validated their expressions using Monte Carlo simulation.
Here, in arbitrary VC geometry, we only expect the first-order scaling expressed in (\ref{eq:RMSrho_R}) to apply, with $H$ interpreted as a measure of VC size.

The easiest derivation of (\ref{eq:RMSrho_R}) builds on discrete-time RW theory.
We know that in boundless 3D space the variance of the distance from the origin, $\langle \br_n^2 \rangle$, scales linearly with the number $n$ of steps taken and with the variance of each step.
Now, the latter is in essence the MFP squared, in this case, equated to $\ell_\rt^2$.
The same scaling applies to all three coordinates individually.
Thus, in a finite space of size $H$, such as the VC, the characteristic number of steps needed to cross that distance is $n_H \sim (H/\ell_\rt)^2$.
To model reflection by the medium, we need to visualize RW in a 3D half-space (rather than fully unbounded 3D space), say $z > 0$.
We then invoke the ``law of first passage'' that gives the probability that a RW that starts in the positive $z$ direction will return to the $z = 0$ plane after exactly $n \ge 1$ steps, which goes as $p_n \propto n^{-3/2}$ \citep{Redner2001}.
Interestingly, the mean value of $n$ is infinite, in view of the very long tail of $p_n$.
However, in finite spaces, the law is truncated at the above characteristic value of $n_H$. 
Therefore, we can compute the mean value of $n$ in reflected light as $\langle n \rangle_R \approx (\sum_1^{n_H} n p_n)/(\sum_1^{n_H} p_n) \sim n_H^{1/2}$, as long as $n_H \gg 1$ (denominator approaches unity).
From there, we can estimate $\rho_{\text{rms},R}^2 \sim \ell_\rt^2 \langle n \rangle_R \sim \ell_\rt^2 n_H^{1/2} \sim \ell_\rt H$, hence (\ref{eq:RMSrho_R}).

The scaling relation in (\ref{eq:RMSrho_R}) was derived with active remote sensing in mind, i.e., a laser beam actually exciting a diffuse reflected radiance field with that RMS radius.\footnote{
This experiment was successfully implemented with real-world stratiform clouds from below \citep{polonsky2005wide} and from above \citep{cahalan2005thor} with, moreover, consideration of laser pulse stretching (i.e., temporal smoothing by the multiple scattering before reflection).}
Equation (\ref{eq:RMSrho_R}) can however be interpreted in passive remote sensing as a cutoff scale in a low-pass spatial filter or smoothing kernel for light reflected off the VC. 
In fact, the expression was first used to interpret quantitatively the phenomenon of radiative smoothing observed as a break in the scaling of wavenumber spectra of LANDSAT images of marine stratocumulus \citep{Davis_etal1997} and reproduced in early Monte Carlo simulations \citep{Marshak_etal1995} where the internal cloud structure was modeled using a turbulence-like stochastic model \citep{Marshak_etal94}.
Due to the spatial heterogeneity in the OS, the light impinging upon the VC is structured, a priori across a wide range of scales.
A fraction $R$ of all that light is reflected (``DC'' component).
Furthermore, it will be reflected back into the OS spatially smoothed: only structures larger than $\rho_{\text{rms},R}$ in (\ref{eq:RMSrho_R}) will be present (``AC'' component).

We can now revisit an aspect of Fig.~\ref{fig:Koch_R_vs_T}a that was overlooked.  
The experiment therein was to increase and decrease the overall opacity of the tutorial 2D Koch cloud model concocted by \cite{Forster_etal21} (e-supplement) and learn from the optically thin-to-thick sequence.
Solar and viewing directions were carefully chosen to show entirely reflected and entirely transmitted radiance fields, and thus support an investigation of \emph{integrated} cloud reflection and transmission inspired by a diffusion-theoretical prediction for how the ratio of those cloud-scale quantities varies with horizontally-averaged COT.
Mean COT was thus boosted by factors of 2 from 2.8 to almost 180.
The interesting smaller-scale phenomenon worthy of belated consideration is that there are five distinct features in the image of the sunlit side of the cloud (in this case, MISR's Ca camera).
Individually numbered in Fig.~\ref{fig:Koch_R_vs_T}a, they are mapped to 1$^\text{st}$- and 2$^\text{nd}$-generation growths in the generation of the Koch fractal in Fig.~\ref{fig:Koch_in+out_VC}a.
The strength of these image features clearly increases with cloud opacity.
This is a clear illustration of how radiative smoothing works.
Extending the application of the radiative smoothing scale for reflected light in (\ref{eq:RMSrho_R}) from the VC to the whole cloud, we can take $H$ = 4~km as the cloud's size.
Dividing $H$ by square-root of $\tau_\rt = (1-g)\tau_\text{avg}$ in Fig.~\ref{fig:Koch_R_vs_T} yields estimates of $\rho_{\text{rms},R}$.
They start at values $\gtrsim H$ for the optically thinnest cases, hence all structures are smoothed.
In sharp contrat, $\rho_{\text{rms},R} \lesssim 1$~km for the most opaque incarnations, hence the resolution of roughly $H/\rho_{\text{rms},R} \gtrsim 4$ equally-spaced structures across the along the sunlit side of the Koch cloud, which projects onto $\approx$36 MISR pixels in the C cameras used here.
If the cloud's opacity was further boosted, ever finer details become resolvable, assuming the camera's pixel size remains somewhat smaller than the finest cloud detail.

Similar questions about radiative smoothing have been asked about light transmitted through optically thick clouds.
\cite{davis2002space} showed that 
\begin{equation}
\label{eq:RMSrho_T}
\rho_{\text{rms},T} \propto H,
\end{equation}
irrespective of $\ell_\rt$ (hence of $\tau_\rt$). 
Again, this is only the first-order scaling, and all that is of interest in the present context. 
The authors derived the exact prefactor and pre-asymptotic corrections in $\ell_\rt/H = 1/\tau_\rt$ for (\ref{eq:RMSrho_T}) in closed-form for plane-parallel slabs, and performed extensive Monte Carlo simulations for numerical validation.
\cite{VonSavigny2002time} further validated observationally the theoretical prediction of \cite{davis2002space} by analyzing across a wide range of scales zenith radiance transmitted to the ground through an extended stratus layer as it was advected across a vertically-pointing narrow-FOV sensor.
The stratus cloud's internal structure is known to have a power-law scaling power spectrum in $k^{-5/3}$ from several km down to scales of a few meters \citep[e.g.,][]{davis1999horizontal}, much less than cloud thickness $H$, which is $\sim$100s of meters.
However, in the observed zenith radiance time-series, once converted into a spatial field using Taylor's frozen turbulence hypothesis, structures less that $\sim$$H$ in scale had been smoothed out.

What does (\ref{eq:RMSrho_T}) tell us about radiation transport across the VC?
Again, a fraction $T = 1-R$ of the sunlight impinging on the VC is transmitted.
We note that, since it was argued that (\ref{eq:RoT_ratio}) applies to the VC, we have $R/T = (H/\ell_\rt)/2\chi$ in the present notations; we can thus solve for $T = 1/[1+(H/\ell_\rt)/2\chi]$, and $R = 1-T$ follows.
So much for the ``DC'' component.
As for the ``AC'' component, it will be smoothed in transmission across a length scale $\rho_{\text{rms},T}$, which is commensurate with $H$, the outer size of the VC.
That is to say that there is really only a DC component: the light transmitted through the VC is spread evenly over the non-illuminated side.

Extending the application of (\ref{eq:RMSrho_T}) from the VC to the whole cloud, we can again look back at Fig.~\ref{fig:Koch_R_vs_T}a, this time focusing on the Cf camera image. 
In this case, the radiance field is dominated by transmitted light for all but the leftmost pixels that associated with the very top of the cloud where the OS is seen directly by both the sun and the camera (cf. Fig.~\ref{fig:Koch_R_vs_T}).
That cloud-top spike in radiance (present at all COTs) is akin to seeing from above a cloud's ``silver lining,'' a forward-scattering phenomenon familiar to any ground observer (scattering angle here is 60$^\circ$).
Otherwise, the image is quite bland at all COTs, in accordance with (\ref{eq:RMSrho_T}) as a cutoff for distinguishable features.

\subsection{Spatial variability in the VC that may matter for CT}

So far, we have treated the VC as if it was a spatially uniform region inside the 3D cloud of interest. 
In real clouds it is of course far from that. 
Are we at risk of biasing future MISR+MODIS CT outcomes if we enforce VC spatial uniformity in the forward model, or as a regularization?
At this stage, we need to further mine the literature for analogous situations and, again, MUSCL comes to mind.
\cite{davis2008multiple} addresses the impact of internal cloud structure on MUSCL spatial and temporal signals, with a concern about what happens to cloud property retrievals if it is neglected.
Being motivated by lidar, only \emph{reflected} light is investigated.
\cite{Davis_etal2009} survey systematically the application of spatial and/or temporal Green functions in cloud remote sensing, both passive and active, and they add new results about the impact of internal cloud structure for light \emph{transmitted} all the way through the cloud.
In both of these studies, the two kinds of internal variability in the 3D cloud extinction field addressed are: (1) cloud-scale vertical gradients, and (2) sub-cloud scale random turbulence.
As far as MUSCL (hence reflection) is concerned, the conclusion is that cloud-scale gradient effects dominate those of the smaller-scale internal turbulence.

\cite{Forster_etal21} state emphatically in connection with their Figs.~3 and 4 that simply replacing a surrogate VC (grey square region in Fig.~\ref{fig:Diffusion_length_Koch}a) by the average extinction therein leads to noticably different cloud images, with up to 20\% changes in pixel-scale radiances.
They therefore proceed in their computational experimentation on the sensitivity MISR data to internal VC structure by replacing the turbulent extinction field inside the surrogate VC with alternative realizations.
These alternatives must however comply with stochastic continuity conditions at the VC/OS interface.
That requires keeping the same cloud-scale vertical gradient across realizations.
Apart from being physically-incorrect, not doing it leads to observable differences in the synthetic MISR imagery.

\begin{figure}
\begin{center}
\includegraphics[width=2.7in]{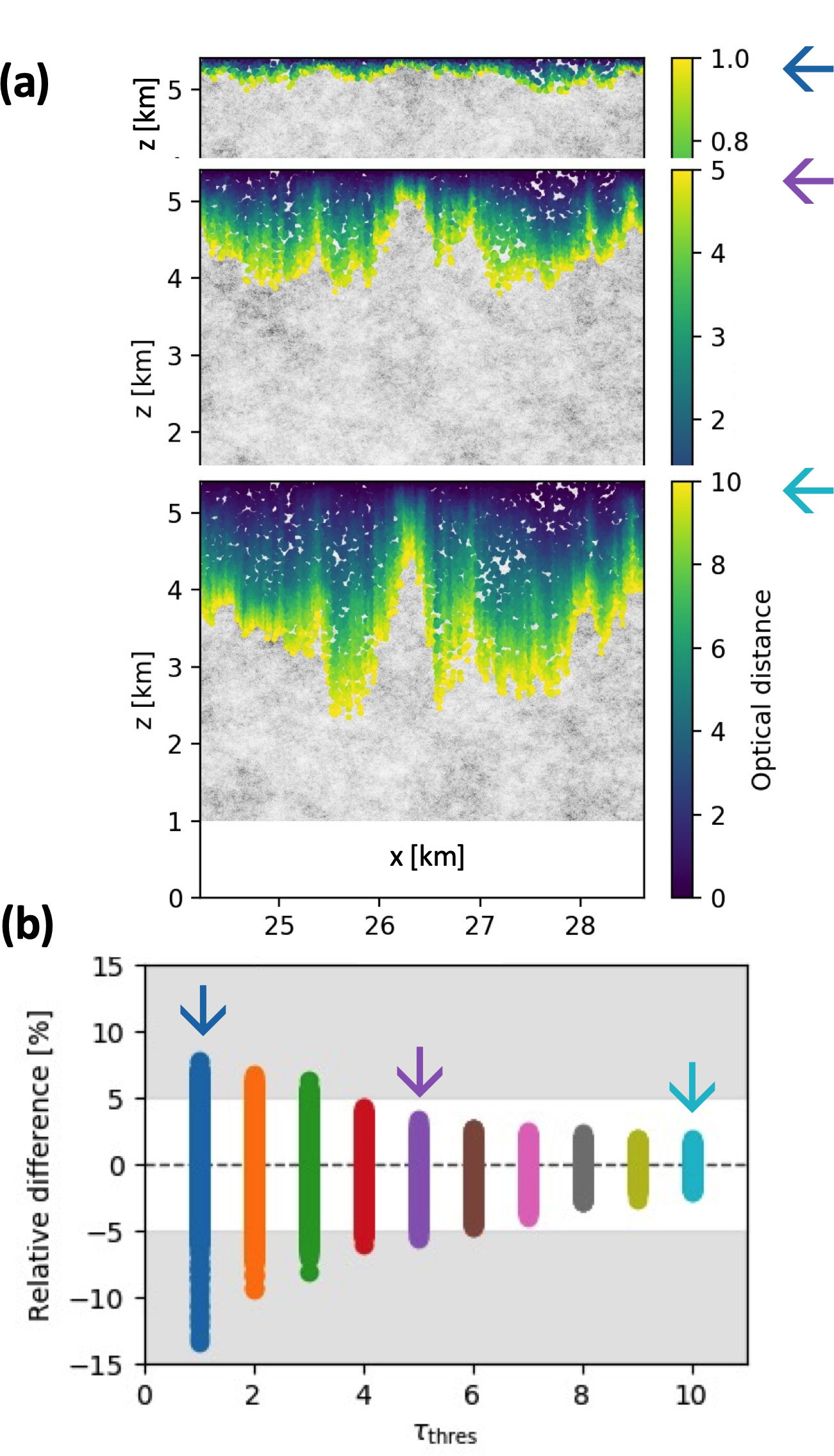}
\end{center}
\caption{
{\bf (a)} A plane-parallel cloud with a nominal COT of 20 is internally randomized with a single realization of a zero-mean 2D scale-invariant stochastic model known as ``fractional Brownian terrain;'' see \cite{Forster_etal21} (e-supplement) for details and references.
Assuming SZA = VZA = 0$^\circ$, the OS/VC interface is determined by setting its optical depth $\tau_\text{thres}$ at integer values from 1 to 10; $\tau_\text{thres}$ = 1, 5, 10 cases are illustrated here.
{\bf (b)} Relative differences in \% cumulated across pixels between the original cloud with 2D spatial variability everywhere and its counterpart where the extinction in the VC (defined as optical depth $>\tau_\text{thres}$) is replaced by its mean value.
We see that, to pass the $\pm 5\%$ bar rationalized by \cite{Forster_etal21}, we need to set $\tau_\text{thres}$ in the 5-to-6 range.
}
\label{fig:VC_turbulence_vs_mean}
\end{figure}

To quantify the impact of multi-scale random turbulence in the VC in the absence of a cloud-scale gradient, we performed Monte Carlo simulations using MYSTIC on a plane-parallel but internally turbulent cloud model.
We ask the defining question of the VC/OS:
{\it At what OS thickness does the detailed internal structure of the VC no longer affect the observed radiance field in a measurable way?}
Figure~\ref{fig:VC_turbulence_vs_mean} displays the outcome where, following \cite{Forster_etal21}, we define ``measurable difference'' as one that exceed 5\% (to be confident that it is not just sensor noise fluctuations).
The pixel-scale Monte Carlo noise is kept at a much lower level.
The VC for a plane-parallel cloud is defined similarly as for the fractal cloud, as shown in Fig.~\ref{fig:VC_turbulence_vs_mean}a, where three selected thresholds are illustrated.
The ``reference'' cloud is 100\% turbulence, while in the OS/VC partitioned clouds the grey region is made uniform at the mean extinction level.\footnote{
We note that the adaptive VC defined here will has, by construction, an extinction discontinuity almost everywhere at its boundary. 
This contrasts with the surrogate VCs, such as the square in Fig.~\ref{fig:Diffusion_length_Koch}a, where continuity is required to bring images differences down to the sensor noise level. 
We attribute this to the fact that not all parts of the square's perimeter are at optical distances in excess of 5 from all the sensors as well as the solar source.
}
Figure~\ref{fig:VC_turbulence_vs_mean}b shows the dispersion of the relative radiance differences across all MISR-scale pixels, and, unsurprisingly, the 5\% tolerance is crossed at $\tau_\text{thres} \approx 5$--6.
There is a systematic negative bias in the radiance differences in Fig.~\ref{fig:VC_turbulence_vs_mean}b, especially for the optically shallow definition of the VC.
That is traceable to the fact that heterogeneous optical media are always less reflective than homogeneous ones for a fixed average extinction in the Independent Pixel Approximation \citep[IPA, e.g.,][]{cahalan94}.
Moreover, in this case of azimuthally symmetric illumination, it was shown that cross-pixel transport deepens the so-called plane-parallel bias \citep{DavisMarshak2001}.

In summary, from the standpoint of CT with MISR and MODIS data, small-scale spatial variability of the 3D extinction field inside the VC is, by definition, in the noise.
However, some large-scale features demand some attention:
({\it i}) at a minimum, there will be a cloud state parameter for the mean extinction in the VC and in the inversion it will likely come out biased low because spatially variable media are always less reflective than uniform ones with the same mean, but there are methods for correcting this bias \citep[e.g.,][]{cahalan94,cairns2000};
({\it ii}) if there is reason to believe that there is a cloud-scale vertical gradient in the VC, for instance in the convective core of a vertically-developing cumulus, then another parameter should be assigned to capture it in the forward model.

\section{OS/VC interaction}
\label{sec:OS_VC_interaction}

We have shown so far that features in images of opaque convective clouds originate by-and-large in extinction field structures in the OS, while the overall brightness contrast between the sunny and shaded sides of the cloud is controlled by the VC.  
We have estimated that roughly 1/3 of the incoming sunlight is reflected back by the OS, and never enters the VC.
Therefore, roughly 2/3 of the sunlight impinging on the cloud does enter the VC, of which a fair fraction is reflected by the VC back into the part of the OS facing the sun.
If the $R/T$ contrast ratio is $\sim$2, which is at the lower end of opaque clouds of interest here (cf. Fig.~\ref{fig:Koch_R_vs_T}), then the VC itself roughly reflects and transmits evenly.
More opaque clouds reflect more, implying that their VCs reflect more than transmit, which is typical of media dominated by diffusive radiation transport.
Here, we examine more closely how the three radiation transport regimes intertwine: streaming (outside the cloud), multiple scattering (in the OS), and diffusion (in the VC).

\subsection{Coupling 3D RT and its diffusion approximation (DA) at the OS/VC interface}

\begin{figure*}[htbp]
\begin{center}
\includegraphics[width=6.5in]{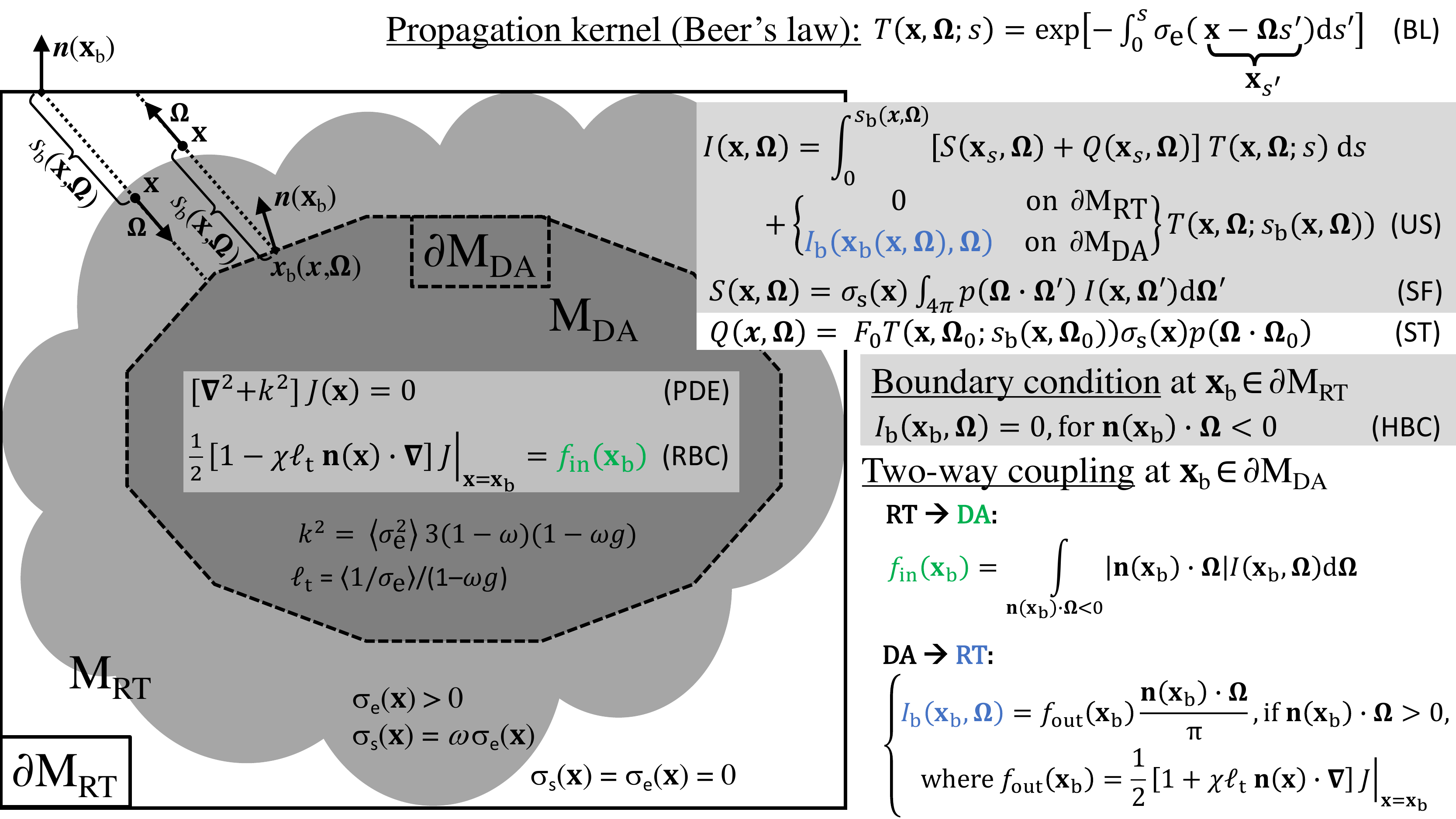}
\end{center}
\caption{Formulation of the computational physics problem where a 3D RTE solver is implemented in the OS and surrounding optical void, and an efficient 3D elliptical PDE solver is implemented in the VC, assumed to be convex.
Special attention is paid on how the two solvers are coupled as radiant energy flows in both directions through the OS/VC interface.
Detailed discussion in the main text.}
\label{fig:hybrid_RT_3D}
\end{figure*}

Figure~\ref{fig:hybrid_RT_3D} sets up the CF forward modeling problem in terms of computational physics.
The complete optical medium is M$_\text{RT}\cup$M$_\text{DA}$, union of an outer region  M$_\text{RT}$ and an inner one M$_\text{DA}$, assumed to be convex.
M$_\text{RT}$ is further divided into a surrounding region of optical void where radiation just streams unobstructed, and the cloud's OS where extinction is non-vanishing, and where transport is therefore modeled by RT with multiple scattering.
M$_\text{DA}$ is the VC and transport therein is accurately modeled with the diffusion approximation (DA).
$\partial$M$_\text{RT}$ is the boundary surrounding the whole computational domain, while $\partial$M$_\text{DA}$ is the OS/VC interface.

The 3D RT problem to be solved in the cloud's OS and surrounding vacuum is defined by the coupled integral equations (US) for an ``upwind sweep'' and (SF) for the ``source \emph{function}'' definition.
That pair of integral equations determines everywhere the diffuse radiance field $I(\bx,\bOmega)$.
The homogeneous (i.e., ``no incoming radiation'') BC for the 3D RT problem on $\partial$M$_\text{RT}$ is expressed in Eq.~(HBC), and it is echoed in (US) as the upper option in the 2$^\text{nd}$ line.
Equations~(BL) and (ST) are the required definitions respectively of the Beer's law-based propagation kernel and of the solar source \emph{term}.
The mechanics of (US) are illustrated with two instances of $(\bx,\bOmega)$ in the upper-left corner.
Note that this key upwind sweep gets far more complicated if the VC is not a convex domain since some beams would then be exiting the VC, then re-entering and re-exiting it, once or more.
In the above, we have shown how an adaptive VC can be constrained to be convex; see Fig.~\ref{fig:adaptive_VC}.

The 3D DA model used inside the VC is defined by Eqs.~(PDE)/(\ref{eq:Helmholtz_PDE}) and (RBC)/(\ref{eq:BCs_DA}), the ``Robin BC,'' where we can set $\chi$ = 2/3.
Thus there are two only parameters in the DA problem to be solved: $k^2$ and $\ell_\rt$, which are expressed in Fig.~\ref{fig:hybrid_RT_3D} in terms of cloud optical properties $(\omega,g)$, themselves dependent on wavelength $\lambda$ and PSD (where $g$ is, however, essentially invariant, cf. Table~\ref{tab:MODIS_SWIR_chans}), and one statistical parameter, $\langle\sigma_\re\rangle$, the VC-averaged extinction coefficient.
Of these quantities, only the latter is a new CT unknown, via $\ell_\rt$ in (RBC) when using MISR data, since $k^2$ in (PDE) vanishes when $\omega$ = 1 at 670~nm.
However, when factoring in the MODIS SWIR wavelengths to gain sensitivity to cloud microphysics, there are three extra unknowns to describe the VC, namely, $\{\omega_{1240},\omega_{1640},\omega_{2130}\}$ via $k_\lambda^2$ in (PDE).
As for MODIS operational retrievals, each SSA translates to a potentially different effective radius $r_\re(\lambda)$ that is representative of an average over the diffusion length-scale $L_\dif = 1/k_\lambda$ in (\ref{eq:Diffusion_length}), as illustrated in Fig.~\ref{fig:Diffusion_length_Koch}.
In a growing convective cloud, we would normally expect $r_\re(1240) < r_\re(1640) < r_\re(2130)$ since $r_\re$ grows with height above cloud base, and $L_\dif$ decreases with increasing $\lambda$.

Finally, in the lower-right corner, we formalize the radiative coupling between the VC/DA zone and the OS/RT zone at their interface $\partial$M$_\text{DA}$.
Specifically, RT$\rightarrow$DA shows how incoming irradiance for the VC is computed from outgoing radiance from the OS.
At the same boundary point, the two expressions in DA$\rightarrow$RT show how outgoing hemispherical flux from the VC becomes an isotropic boundary source for the OS.

To the best of our knowledge no 3D RT solver has implemented this hybridization, and it would be a good idea to do so.
In principle, it should greatly accelerate the solution of the of the forward model for multi-angle imaging signals with tolerable loss of accuracy, commensurate with sensor noise.
Indeed, there are very efficient methods for solving the boundary-value PDE problem in (PDE)--(RBC), e.g., using sparse matrix inversion.\footnote{
We note here that, in numerical analysis, typically much effort (and CPU time) is spent on ensuring very high accuracy of the PDE solution, possibly verging on machine precision. 
Such ultra-high numerical accuracy may be important for problems where the Helmholtz equation describes the physics exactly, e.g., in electrostatics or electrodynamics. 
However, in the present context, it is patently an approximation of the true physics, which is RT. 
Therefore a faster lower-precision solver is desirable.}
From the CT inverse problem standpoint, we also anticipate a major acceleration since all the values of extinction on the 3D grid inside the VC are replaced by just two unknowns (assuming, for simplicity, a uniform VC and prescribed microphysics).
Thus, as candidate clouds for CT-based reconstruction get bigger, the number of unknowns increases as the cloud's surface rather than its volume.

\subsection{Hybrid forward RT modeling illustrated in uniform plane-parallel media}

\begin{figure}[htbp]
\begin{center}
\includegraphics[width=2.7in]{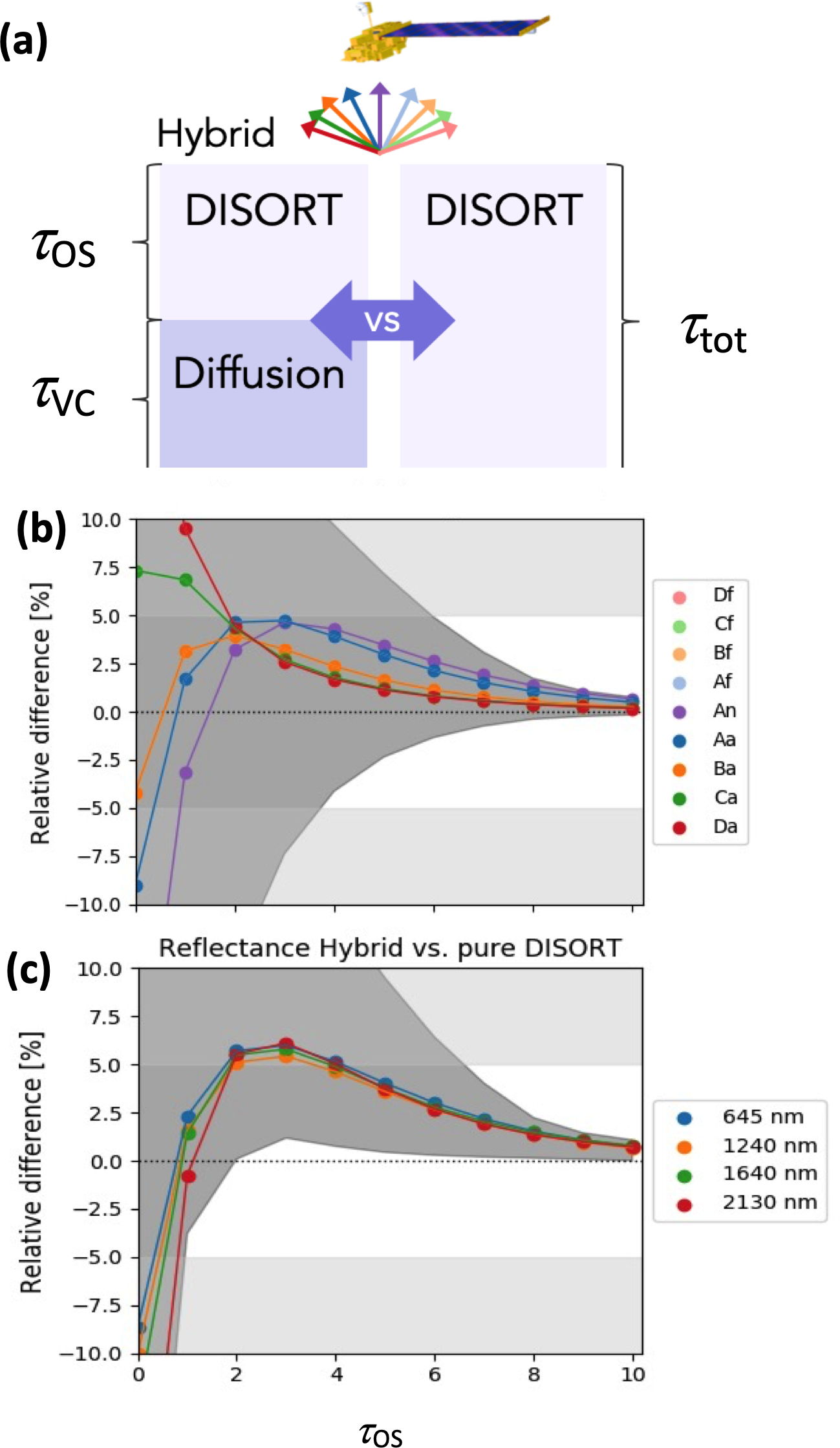}
\end{center}
\caption{
(a) Schematic of the hybrid 1D RT model: DISORT on the top (in OS); diffusion model at the bottom (in VC).
(b) Performance of the hybrid model w.r.t. DISORT benchmark for all nine MISR view angles.
(c) Same as (b) but for all four MODIS spectral channels of interest here (red and 3$\times$SWIR).
If $\tau_\text{OS}$ = 0, the model is pure diffusion; if $\tau_\text{OS}$ = 10, the model is pure DISORT but only in the special case where $\tau_\text{tot}$ = 10 too.
Colored curves are for a representative configuration: $\tau_\text{tot}$ = 20; $r_\re$ = 10~$\mu$m; and sun at zenith (hence same answers for MISR's fore- and aft-views).
Grey regions are the spread in model error for the whole parameter set described in the main text.
}
\label{fig:hybrid_RT_1D}
\end{figure}

Implementation of the hybrid 3D RT forward model described in Fig.~\ref{fig:hybrid_RT_3D} is out of scope for the present study.
However, a limited test of the basic idea can be easily performed in 1D RT; see Fig.~\ref{fig:hybrid_RT_1D}a.
We only need to have access to an accurate and flexible 1D RT solver such as DISORT \citep{Stamnes_etal_1988} to produce benchmark results and play the role of the RT model in the OS.
We will also need two closed-form expressions from diffusion theory to emulate the computational solution of the diffusion problem in the VC.

One expression is the albedo $R_g(\tau_\text{VC},\mu_0) = 1-T_g(\tau_\text{VC},\mu_0)$ from (\ref{eq:T_deltaEdd}), itself coming from the $\delta$-Eddington version of two-stream theory for non-absorbing (a.k.a. conservative) plane-parallel optical media.
The optical thicknesses of the VC and OS are denoted $\tau_\text{VC}$ and $\tau_\text{OS}$, respectively. 
The other expression is $R_\text{dif}(\tau_\rt/2\chi) = 1/(1+2\chi/\tau_\rt)$, where $\tau_\rt = (1-g)\tau_\text{VC}$ is the scaled optical thickness of the VC, and $\chi$ can be set to 2/3.
In contrast with $R_g(\tau_\text{VC},\mu_0)$, which models a collimated solar beam, this last expression applies to diffuse illumination at the upper boundary.
It results directly from (\ref{eq:RoT_ratio}) for the $R/T$ ratio in the diffusion limit, which applies to plane-parallel media \citep{davis2002} combined with radiant energy conservation, i.e., $R + T = 1$.
DISORT is configured for a collimated solar beam (SZA = $\cos^{-1}\mu_0$) at the upper boundary, and a Lambertian reflector at the lower boundary with an effective albedo 
\begin{eqnarray}
\alpha &=& f_\text{dir} \, R_g(\tau_\text{VC},\mu_0) \nonumber \\
       &+& (1-f_\text{dir}) \, R_\text{dif}\left( (1-g)\tau_\text{VC}/2\chi \right), 
\label{eq:hybrid_RT_1D}
\end{eqnarray}
where $f_\text{dir} = \exp(-\tau_\text{OS}/\mu_0)$.
DISORT uses Mie scattering phase functions similar to that shown in Fig.~\ref{fig:PF_convolve} (with an asymmetry factor $g$ = 0.86), for effective droplet radii $r_\re$ = 5, 10, and 15 $\mu$m and wavelengths specified below.
For MODIS SWIR wavelengths, an expression similar to (\ref{eq:hybrid_RT_1D}) is used at the lower boundary, but where the closed-form diffusion theoretical models have an extra argument, namely, SSA $\omega$ from Table~\ref{tab:MODIS_SWIR_chans}.
We refer to \cite{MeadorWeaver80} or \cite{Davis_etal2009} for formulas for $R_{\{\omega,g\}}(\tau_\text{VC},\mu_0)$ and $R_\text{dif}\left( (1-g)\tau_\text{VC}/2\chi, L_\dif/\ell \right)$, where the 2$^\text{nd}$ non-dimensional argument is found in (\ref{eq:Diffusion_length}) and Table~\ref{tab:MODIS_SWIR_chans}.
Finally, DISORT is run for a single layer of optical depth $\tau_\text{OS} = \tau_\text{tot} - \tau_\text{VC}$, where $\tau_\text{tot}$ is set to 10, 20, and 40.
Accuracy benchmark results are obtained by setting $\tau_\text{VC}$ = 0, hence $\tau_\text{OS} = \tau_\text{tot}$ (pure DISORT for each case).

Figures~\ref{fig:hybrid_RT_1D}bc display prediction errors of the above 1D RT hybrid model for given $\tau_\text{OS}$ between 0 and 10, and a broad range of parameters: $\tau_\text{tot} \in \{10, 20, 40\}$; $r_\re \in \{5, 10, 15\}$  [$\mu$m]; and SZA = \{0,60\} [$\null^\circ$].
In addition, we scan VZA across MISR's nine viewing directions (cf. Fig.~\ref{fig:Koch_in+out_VC}) for its red channel ($\lambda$ = 670~nm) in panel (b).
In panel (c), VZA = 9$^\circ$ and four MODIS spectral channels are sampled ($\lambda$ = 645, 1240, 1640, and 2130 [nm]), with the three latter SWIR channels having significant liquid water absorption (cf. Fig.~\ref{fig:Diffusion_length_Koch}) and, from there, cloud microphysical sensitivity to be exploited.
In Figs.~\ref{fig:hybrid_RT_1D}bc, one representative case ($\tau_\text{tot}$ = 20, $r_\re$ = 10~$\mu$m, SZA = 0$^\circ$) is highlighted; error over the 9 MISR views (same answer for fore-and aft-VZAs when SZA = 0$^\circ$) and the 4 MODIS wavelengths are plotted in different colors. 
The full spread of the outcomes across parameter space is captured with the grey region around those selected means.
We see that the maximal error for all MISR views becomes less than the highlighted $\pm$5\% when $\tau_\text{OS} \gtrsim 6$, while the counterpart for MODIS across all spectral channels crosses that max-error threshold at $\tau_\text{OS} \gtrsim 6.5$.

In essence, Fig.~\ref{fig:hybrid_RT_1D} does for angular and spectral diversity what Fig.~\ref{fig:VC_turbulence_vs_mean} does for spatial variability across pixels.
Interestingly, the two experiments independently put the optical depth of the OS in the 5-to-6 range, yet again.
Both studies transpose the notions of OS and VC from vertically-developed clouds to a plane-parallel setting, reminding us that the OS is fundamentally the \emph{radiative boundary layer} of the optical medium, irrespective of shape and internal structure.
Across this boundary layer, radiation transport transitions smoothly from a diffusion process in the VC to the streaming that occurs outside the cloud and onto the solar source and the overhead sensors.
This transition has to be modeled with bone fide 3D RT, but the transport of solar radiation across the VC can be vastly simplified in the interest of computational efficiency.
We have advocated a standard 3D diffusion model in this paper, but it is not the only option.
For instance, \emph{generalized} radiative transfer theory \citep{davis2014generalized,xu2016markov} and non-classical linear transport \citep{LarsenVasques11,frank2010generalized} are related but distinct homogenizations of 3D RT in spatially variable optical media, and the latter has its own diffusion limit \citep{frank2015nonclassical}.

\section{Summary and discussion of our findings}
\label{sec:discuss}

\cite{Forster_etal21} discovered the VC of opaque convective clouds by using synthetic multi-angle imagery of a toy model to search for that part of the cloud that paradoxically does nothing measurable for the image of the cloud, at least not the finer details of the cloud image (as observed by realistic sensors with $\approx$3\% radiometric noise levels).
Those finer details are not only what makes clouds fascinating phenomena to contemplate and ponder.
They are also the image ``features'' that enable automated retrievals of cloud top heights and winds using stereoscopic methods to process satellite imagery, e.g., from MISR.
Moreover, the physical linkage between spatial structures across multi-angle/multi-spectral imaging of vertically-developed cumulus clouds is key to their 3D imaging using optical tomographic methods.
That is indeed the challenge of CT, a promising new technique in passive cloud remote sensing in the VNIR-SWIR, which has already motivated a multi-nanosat mission, CloudCT, currently under development in Germany and Israel \citep{cloudct}.
CloudCT will use \emph{high} spatial resolution cameras covering a narrow swath to build directly on recent CT demos \citep{Levis_etal15,Levis_etal17,Levis_etal20} using LES clouds and AirMSPI data \citep{diner2013}, where both grid-scales and pixels are $\sim$20~m.

Here, we pursue CT using existing space-based imager data, specifically, with MISR and MODIS on Terra, which have relatively modest spatial resolutions, between 250 and 500~m pixels.
This endeavor is not just about legacy data. 
The next generation of NASA's Earth observing assets in space will have comparable or improved capabilities; the Multi-Angle Imager for Aerosols investigation \citep[MAIA,][]{diner2018advances} comes to mind, where then main improvement for cloud remote sensing is to have access to microphysics through both polarization and (now multi-angle) SWIR.
At any rate, graduating CT from airborne to satellite sensing is a quantum leap in computational methodology for both forward and inverse problems because there will be strong sub-pixel spatial variability to reckon with and, moreover, pixel-scale optical thickness can far exceed unity.
We recall that efficient (hence deterministic) numerical solvers for 3D RT assume uniformity and mandate at least semi-transparency inside each grid-cell.
We are convinced that discovery of the VC will be key to meeting this new forward-modeling challenge in computational 3D RT, as well as in the proper definition of the large inverse problem in CT using a combination of MISR's multi-angle and MODIS's multi-spectral data.

Having found in the VC what does \emph{not} affect the intricate texture of cloud imagery, we set out in the present study to understand on the basis of physics what does encode 3D cloud structure into 2D imagery, respectively, output and input of CT.
Here are our main findings along the way:
\begin{enumerate}
\item 
We established that, whatever structure there is in the VC, it will not influence the fine-scale structure of the cloud images in a detectable way.
However, {\bf the overall optical ``mass'' (mean extinction and size) of the VC does control the cloud-scale radiance contrast} between the part of the cloud's outer boundary that is directly illuminated (reflected light) and its self-shaded side (transmitted light).
    \begin{itemize}
    \item Moreover, there is a near-linear relationship between the reflected/transmitted radiance contrast ratio and the mean optical thickness of the whole 3D cloud---a fact that will prove helpful in the initialization of the CT optimization procedure.
    \end{itemize}
\item 
We showed that the {\bf 3D RT modeling in the VC can be vastly simplified} by assuming that it is uniform, with the possible exception of a cloud-scale vertical gradient, which either way reduces significantly the number of unknowns in the CT inverse problem.
    \begin{itemize}
    \item In addition, the 3D RT in the VC can be modeled with sufficient accuracy by using the 3D diffusion approximation, for which there exist extremely fast numerical solvers; this is good news for the forward model embedded in the CT optimization problem.
    \item The RT physics in the VC is therefore analogous to a steady-state diffusion process where RWs start at the illuminated side of the VC and can end anywhere but, if starting on the illuminated side, generally not so far from where they started.
    \item RW theory was used to derive the scaling of the radiative smoothing scale as the harmonic mean of the only two lengths in the diffusion problem in the absence of absorption: size of the medium and size of the steps, a.k.a. mean-free-path (MFP) but, in this case, the ``transport'' MFP, which is where, on average, an effectively isotropic scattering happens (factoring in the forward scattering).
    \end{itemize}
\item 
Outside the VC, we find the cloud's OS, which \cite{Forster_etal21} would have defined as the first 5-to-6 optical depths of the medium, coming in from the solar source or any of the imaging sensors, but before reaching the VC.
Forster and coworkers found that 5-to-6 optical depth for the OS/VC interface empirically, by requiring that the impact of any physically reasonable rearrangement of the extinction field inside the VC on the observed imagery does not exceed the sensor noise threshold for a radiance difference.
Here, {\bf we view the OS as the radiative analog of the boundary layer in fluid dynamics}---the region where the presence of the boundary strongly influences the structure of the radiance field.
    \begin{itemize}
    \item We confirmed numerically the 5-to-6 optical depth of the OS, but coming from the opposite direction as Forster and coworkers; we ask: At what optical depth into the cloud does an opaque object cease to leave a detectable imprint in the remotely-sensed cloud image?
    \item Furthermore, we use Green function theory for 3D RT in an infinite medium to predict the 5-to-6 optical thickness of the OS based only on phase function characteristics (indeed only its asymmetry factor) by recasting 3D RT as a RW on the 2D unit sphere of directions and an associated RW in 3D space, but with angularly-correlated steps.
    \item Steps in the RW on the 2D sphere are isotropic, but the fact that they are relatively small compared to the radius of curvature strongly impacts the related RW in 3D space, where step lengths are independent, but directions are not; this leads to a finite drift of the RW position in the direction of the 1st step as well as to a lateral dispersion.
    \item Both longitudinal drift and lateral dispersion are computed analytically and numerically in unbounded 3D space and half-space for every step; in the limit of an infinite number of steps, the drift is identical to the ``transport'' or scaled MFP.
    \item Three other experiments independently confirm the same 5-to-6 prediction: two were conducted herein numerically (in pixel- and angle-space, respectively) while the third path was to relate our theoretical and computational results to a controlled laboratory investigation by \cite{Bohren_etal95} on how optically thick a cloud needs to be to totally obscure the sun.
    \end{itemize}
\item 
Finally, motivated by securing for CT highly-desirable sensitivity to cloud particle size, we examined how cloud images are impacted by absorption in liquid water droplets at three SWIR wavelengths sampled by MODIS. 
Based on the diagnostic diffusion length scale, we find that for moderately opaque cumulus clouds (mean optical thickness $\sim$22), {\bf the effective microphysical probing depth is commensurate with cloud size at 1240~nm, reaches into the VC at 1640~nm, but barely covers the OS at 2130~nm}; see Fig.~\ref{fig:Diffusion_length_Koch}.
\end{enumerate}
The last item will be critical in the study of cumuliform clouds in convective dynamical regimes using CT to recover information on the droplet size profile.

At a fundamental level, VNIR-SWIR cloud images are formed by two intertwined radiative diffusion processes. 
\begin{itemize}
\item One unfolds in the OS: 
{\bf a RW on the 2D unit sphere that models the gradual loss of directional memory} in the sunlight as it penetrates the cloud, and that drives a spatial RW with strongly correlated steps in direction space (due to the forward-peaked phase function). 
\item The other unfolds in the VC: 
{\bf a standard RW in 3D space akin to particles in Brownian motion} that start at the OS/VC interface facing the sun and end anywhere on it, but generally not very far to the starting point if it is a reflection rather than a transmission event.
\end{itemize}
We reckoned that $\sim$1/3 of the red-channel sunlight impinging on a cloud opaque enough to have a sizable VC never reaches it, as it escapes back to space before crossing the OS/VC interface.
In the absence of absorption (i.e., MISR- and MODIS-red channels), the VC partitions the remaining $\sim$2/3 of the incoming light between reflection and transmission, using the solar direction as a discriminator, with the former dominating the latter more-and-more as the VC increases in opacity.\footnote{
``Sideways'' deflection by the VC is a concept that literally borrows from the two main categories of reflected and transmitted light in a way that is essentially subjective.}

Finally, the sunlight escapes the cloud---with or without reaching the VC---streaming into directions defined by individual pixels at the focal plane of the space-based imaging sensor.
Since it is (still or back) in the OS, this outgoing light is again following a RW of the first kind---on the 2D sphere, driving an angularly-correlated one in 3D space.
However, {\bf this last RW is best visualized as starting at the sensor's pixel and sending well-collimated ``reciprocal'' or ``adjoint'' light into the cloud}.
Therein this backtracking light diffuses across direction space and propagates in 3D space until it crosses paths \emph{at every step} with forward-propagating sunlight, either in the OS or at the OS/VC interface.
At those points, all physically realizable paths from the sun to the sensor through the cloud have thus been accounted for.\footnote{
Certain Monte Carlo 3D RT codes used in photorealistic computer graphics implement literally this forward-and-backward methodology, for instance, using the powerful ``photon mapping'' approach \citep{jensen2009realistic}.}

\section{Case study}
\label{sec:Case_study}

\begin{figure}[htbp]
\begin{center}
\includegraphics[width=3.14in]{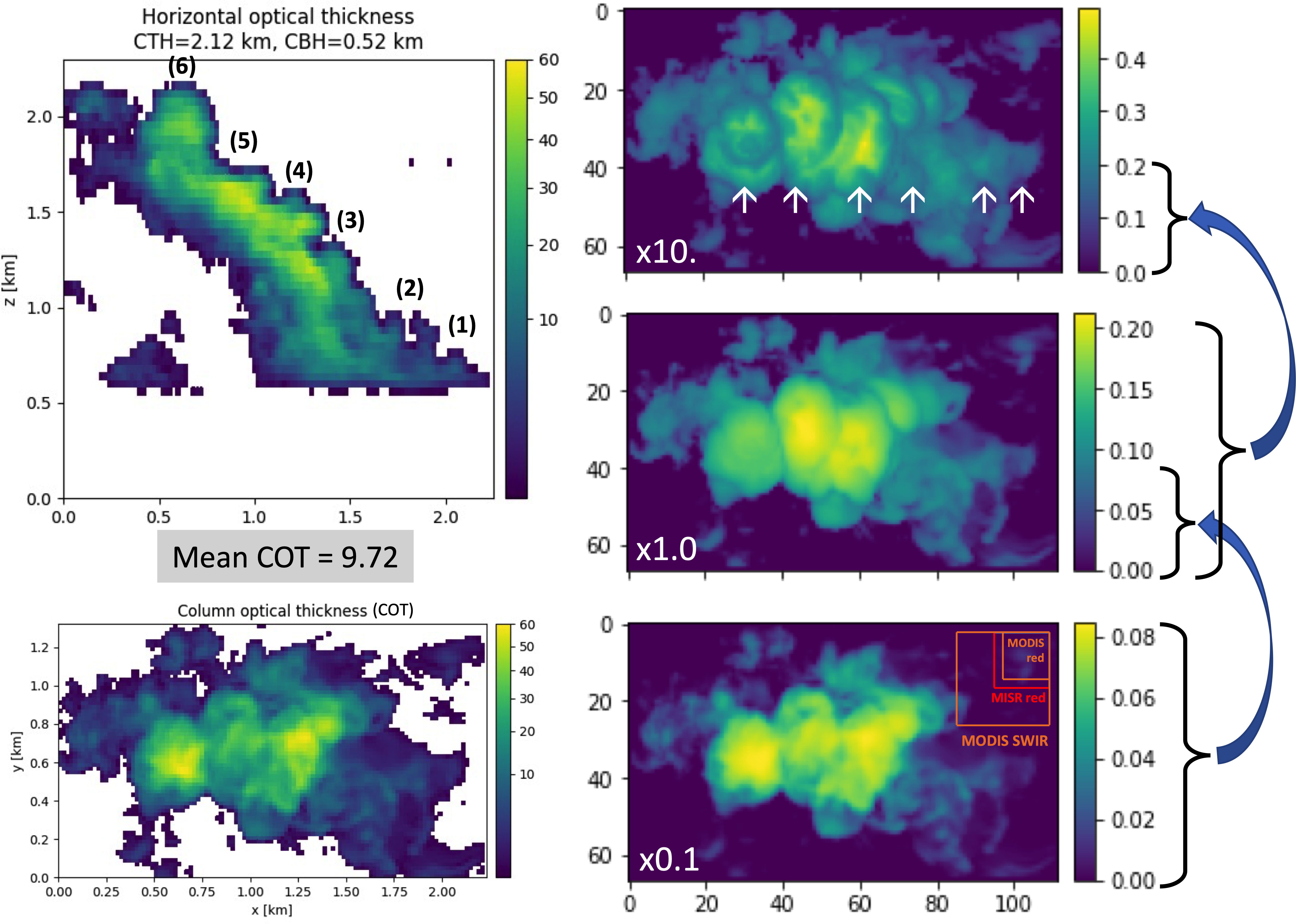}
\end{center}
\caption{
{\bf Left panels: }
A single vertically-developed cloud generated by the JPL LES \citep{matheou2014large} viewed as two zero-masked optical thickness fields plotted on a log-scale: 
lower panel is the usual vertical integration; 
upper panel shows integrals of the extinction coefficient along the $y$-axis.
An effective radius of 10~$\mu$m was assumed to convert the gridded LWC field into a 3D distribution of extinction values for $\lambda$ = 670~nm (MISR red); the mean COT across cloudy columns is then 9.72.
{\bf Right panels: } 
The middle panel shows a synthetic image of the cloud at the native LES resolution computed with the popular open-source 3D RT solver SHDOM \citep{evans1998spherical}.
Solar and view angles are both set to 0$^\circ$, hence no shadows are visible, no cloud surface hidden (except by overhangs); surface is black; no aerosol nor Rayleigh scattering is considered.
Top and bottom panels are images of the same cloud, but with extinction everywhere multiplied or divided by 10, respectively, bringing the mean COT to 97.2 (top) and 0.972 (bottom).
Radiometric scales are more than doubled at each step to accommodate the increasing range of radiance values from bottom to top.
For context, the footprints of MISR-red, MODIS-red and -SWIR pixels are shown as insets in the bottom panel.
Numbers and pointers in the top two panels are discussed in the main text.
}
\label{fig:LES_big_SHDOM}
\end{figure}

To close out the present study, we offer Fig.~\ref{fig:LES_big_SHDOM} where we revisit the protocol used in Fig.~\ref{fig:Koch_R_vs_T} in which cloud opacity is iteratively boosted.
However, we start here at an overall smaller optical thickness ($\sim$1 on average, down from $\sim$3) and we use two factors of 10 rather than six factors of 2.
Rather than our ``Koch fractal'' toy model, this time we use a moderate-size 3D cloud from an LES simulation using RICO \citep{rauber2007rain} forcing conditions (left panels), and the 3D RT is performed using deterministic rather than Monte Carlo code to generate the 3 nadir images (right panels).
The image pixel size is the same as the LES horizontal grid scale, namely, 20~m; for the present context, we also show (lower right) the areas covered by MISR-red, MODIS-red and MODIS-SWIR pixels, respectively, 275, 250, and 500~m, from Fig.~\ref{fig:EUREC4A}.
That makes clear the challenge of tomographic reconstruction of the 3D cloud structure from MISR and MODIS data.
This is indeed one of the two LES clouds successfully reconstructed by \cite{Levis_etal15} at the native LES resolution (similar to what can be achieved with airborne sensors)\footnote{
Such resolutions can of course be achieved from space, but at the expense of image swath and SNR.}
where the pixels and voxels are optically thin and internally uniform, as required for deterministic 3D RT modeling.
However, the two Terra-based imagers are plagued with pixels that are potentially opaque and certainly heterogeneous, in blatant contradiction with said requirements.

We therefore set out to help overcome these challenges by improving our fundamental understanding of how cloud images are formed in the first place.
How can this advance help to comprehend the three images to the right in Fig.~\ref{fig:LES_big_SHDOM}?
We first notice that the top and bottom images show the sharpest features, but we will argue that it is for opposite reasons.
\begin{itemize}
\item 
We start with the optically thinner (bottom) case where the maximum optical thickness is $\approx$6 so, in essence, this cloud is all OS.
The MFP (corresponding to one unit of optical thickness) in fact exceeds cloud thickness in all of the many pixels near the meandering cloud edges.
These parts of the image are thus dominated by single-scattered light, which never leaves the column it first hits when both sun and sensor are at zenith, as is the case here.
In the most optically thick columns around the center, there is however clear evidence of multiple scattering.
That said, the strongly forward-peaked phase function keeps, on average, the light relatively close to the column it first enters, until it is finally backscattered.
In short, this image is almost like an X-ray of the cloud if one compares it with the vertically-integrated extinction map to its right, plotted on a log scale.
\item 
The middle panel shows the cloud at its ``natural'' opacity, where the average cloudy column has a respectable optical thickness of nearly 10 and, in the center of the cloud, we find optical thicknesses in the many 10s.
This cloud therefore has a well-developed VC, although large portions of the cloud with low COT will not be part of this VC and, moreover, it will probably not be itself optically thick enough to use the diffusion approximation. 
Nonetheless, since there is a VC, we can invoke the above-mentioned idea of a radiative smoothing scale, which quantifies to first order the lateral distance between entry and escape of the sunlight.
Factoring in the $\approx$45$^\circ$ inclination of this cloud (upper left panel), its physical thickness along the diagonal is on the order of 0.6~km and, in the central region, its optical thickness along the vertical is now in the 30--60 range, according to the legend (lower left panel), hence 20--40 projected along the diagonal.
With these rough estimates, our Eq.~(\ref{eq:RMSrho_R}) for the reflective smoothing scale yields 0.25--0.35~km, hence 0.17--0.25~km projected back onto the horizontal plane.
To the eye, that looks like a reasonable estimate of the resolvable level detail in the optically thick center of the cloud, in comparison with the optical thickness map (lower left).
\item 
In the optically thicker (top) case, the maximum optical thickness is now $\approx$600.
As evidenced by the logarithmic vertical optical thickness scale, all but the most peripheral columns have COT in excess of 100, so this cloud is practically all VC, thus making radiative smoothing theory more accurate.
According the reflective smoothing scale in (\ref{eq:RMSrho_R}), all lateral transport distances in the middle panel are reduced by a factor of $\sqrt{10}$.
Therefore, we should be able to resolve details between 55 and 80~m in size, about 3 or 4 pixels, respectively, which to the eye seems about right.
We note that LES cloud structure is not expected to be realistic at these small scales, but somewhat too smooth due to the nature of the numerics in the fluid mechanics.
This deficit in small-scale variability is in addition to the radiative smoothing, and explains the ``blobby'' appearance of the cloud in the simulated image.
Interestingly, and in sharp contrast with the lower-right image, there is only a vague resemblance here between the COT map (lower left) and this simulated cloud image.
That is because it is clearly the variable height of the cloud's upper surface that shapes this image.
One can count 6 ``levels'' leading from cloud base to the top of the highest convective ``plume,'' as highlighted in the two top panels.
\end{itemize}

\section{Conclusion \& outlook}
\label{sec:concl}

We set out to uncover the physical principles at work in the formation of convective cloud images, as observed from space by MISR and MODIS on Terra, and kindred imaging sensors operating in the solar spectrum, present and future.
We were motivated by the challenges that arise, due to pixel size and spatial variability, when migrating 3D cloud tomography (CT) algorithms from high to moderate spatial resolution (i.e., airborne to satellite platforms).
We did not resolve these challenges, but we are now in a better position to do so, building on a new understanding of cloud image formation.
That said, the present study applies not only to CT but, broadly-speaking, to any remote sensing operation that uses cloud image ``features,'' e.g., stereographic determination of cloud-top heights and winds.

We have indeed reduced the complex radiation-cloud interactions that culminate in the formation of an image to a single unifying concept in statistical physics, namely, diffusion processes, equivalently, random walk (RW) theory.\footnote{
Not to be confused with the so-called ``Markov chain'' formulation of computational radiative transfer that is closely related to RW theory in ``transport'' space, a merger of 1D or 3D space of positions and the 2D space of directions.
See \cite{esposito1978radiative} and \cite{xu2011markov} for implementations in plane-parallel media, without and with polarization, respectively.}
Furthermore, we have identified a three-part storyline where RW theory plays a central role and that unfolds as sunlight travels from the illuminated cloud boundary to the imaging sensors in space.
\begin{itemize}
\item 
The {\it First Act} is about how the spatially uniform and unidirectional incoming solar beam gradually loses its directionality over the first 5-to-6 optical depths into the cloud; uniformity is of course also broken as the sunlight flows in and around the spatially variable extinction field.
In this first segment, light travels across the region we call the illuminated section of the ``outer shell'' (OS).
The deep RW here is on the 2D unit sphere of directions that, in turn, drives a \emph{non-standard} RW in 3D space with directionally-correlated steps.
The two consequences of these angular correlations are that: 
({\it i})  the mean position of the RW drifts systematically in the direction of the incoming solar illumination; 
({\it ii}) simultaneously, light coming in at a specific point on the cloud boundary spreads laterally as it penetrates the medium.
This first diffusion process ends when the light is well on its way to ``forgetting'' its original direction of propagation.
\item 
As argued next, {\it Act Two} is in fact optional, but only under some special circumstances. 
It starts whenever the now very diffuse sunlight reaches the ``veiled core'' (VC) of the cloud, as defined empirically in our previous study \citep{Forster_etal21}; therein sunlight executes a \emph{standard} (isotropic step direction) RW in the available 3D space, which is necessarily finite in vertically-developed convective clouds.
Steps in this RW are isotropic because the extinction field has been scaled back to account for the forward-peaked phase function following, e.g., the similarity theory of \cite{joseph76delta}.
Based on RW theory, we can compute how far the light travels on average along the VC boundary from entry to escape, and relate this lateral radiation transport to the phenomenon of radiative smoothing \citep{Marshak_etal1995}.
We can also compute from RW/diffusion theory the net radiative flux across the VC that, in turn, determines the cloud-scale contrast in radiance between the brighter sunlit side of the cloud to its dimmer self-shaded side.
\item 
The {\it Final Act} is like the first, but in reverse since we are now interested in how ``reciprocal'' or ``adjoint'' light that starts at a specific image pixel, i.e., a small area in 3D space in combination with a specific direction on the 2D sphere.
By reciprocity, this backward propagation from the sensor into the cloud is affected by the same angular smearing and spatial spreading as the incoming solar beam.
Spatially, this last RW quantifies how, in principle, the entire cloudy medium can affect the radiance in a single pixel/angle.
In fact, the intensity of this adjoint light has been called ``importance,'' following \cite{Marchuk1964}. 
Where it is high, the ambient light coming from the solar source, directly or not, contributes a lot to the pixel-scale radiance and, conversely, where it is low, the ambient light contributes little.

\quad Depending on the pixel and viewing direction, there can be strong overlap between the forward and backward propagating light.
If that happens in the (illuminated part of the) OS, typically at relatively low orders-of-scattering, then it is an opportunity for the sunlight to flow from entry to escape without crossing the OS/VC interface, thus making the Second Act unnecessary.
This shortened playbook is conducive to generating strong image features, and we reckoned that it occurs for $\sim$1/3 of the incoming sunlight.

\quad On top of the $\sim$2/3 that reaches the (illuminated side of) the VC, there are always MISR viewing angles where many if not all pixels have no such overlap with incoming light.
Either way, there is a significant chance that the adjoint light reaches the OS/VC interface.
In that case, the highly diffuse outgoing light field produced across the OS/VC interface during Act Two is what the back-propagating adjoint light collects for the pixel/direction of interest.
In this case, cloud structures in OS, as seen from space are essentially being backlit with the relatively dim and diffuse light that has filtered through the VC.
Image features here will therefore not be as sharp as on the sunlit side of the cloud where backscattered light that never entered the VC prevails.
\end{itemize}
3D cloud structure is thus encoded in MISR's multi-angle imagery using the rules of 3D radiative transfer (RT), and the goal of CT is to invert that map.
However, to do this with the larger pixels and larger clouds, we need to reformulate the CT problem based on insights from physics (here) and computational experimentation \citep{Forster_etal21}, bearing in mind the finite radiometric precision of our sensors.
Forster and coworkers defined a cloud's VC by locating the in-cloud region where the impact of physically-plausible structural changes inside the VC are not detectable, i.e., result in radiance changes at best commensurate with the noise level.
They thereby established that the VC is the in-cloud region where only mean extinction should be a CT target, thus drastically reducing the number of unknowns in the inverse CT problem and avoiding a situation leading to overfitting.
Here, we furthermore suggest that, into the future, the forward 3D RT model in the CT inversion scheme can be vastly simplified in the VC by using the diffusion approximation.

MISR's red channel provides the multi-angle imagery required for CT at 275~m resolution, and we have so far used prescribed uniform cloud droplet sizes for our sensitivity studies.
However, for actual CT, experience tells us that we want to do better than that, and we therefore plan to include in the CT optimization MODIS's SWIR channels where liquid water absorbs in proportion to the droplet's volume.
Even if there is just one viewing direction, we thus gain sensitivity to cloud microphysics, at least at the cloud top and edges.
Implementation details (e.g., prescribed or prior profile of effective radius) are a subject for future study, but we showed here already that, at SWIR wavelengths, the VC will be larger than for non-absorbing VNIR wavelengths.
That is more good news for future forward 3D RT modeling at the core of CT inversion algorithms.

%





%
%

%
\acknowledgments
The research was carried out at the Jet Propulsion Laboratory, California Institute of Technology, under a contract with the National Aeronautics and Space Administration (80NM0018D0004).
LF received funding from the European Union's Framework Programme for Research and Innovation Horizon 2020 (2014-2020) under the Marie Sk\l{}odowska-Curie Grant Agreement No. 754388 and from LMU Munich's Institutional Strategy LMUexcellent within the framework of the German Excellence Initiative (No. ZUK22).
AD was funded primarily by NASA's SMD/ESD Radiation Sciences Program under the ROSES TASNPP element (contract \#17-TASNPP17-0165). 
He also acknowledges a strong influence on his thinking from the ``Planetary Boundary Layer'' working group at JPL supported by the SRTD program.
We are thankful for many fruitful conversations on cloud CT with Aviad Levis, Katie Bouman, Masada Tzemach, Yoav Schechner, Jesse Loveridge, and Larry Di Girolamo.

\bibliographystyle{ametsoc2014}
\bibliography{VC_physics.bib}

\vspace{24pt}
\copyright 2021 California Institute of Technology. Government sponsorship acknowledged.

\end{document}